\documentclass[12pt]{article}
\usepackage{amsmath,amsfonts,amssymb,amsxtra}

\makeatletter\@addtoreset{equation}{section}\makeatother

\newcommand{\preprint}[1]{\begin{table}[t]  %%
             \begin{flushright}               %%
             {#1}                             %%
             \end{flushright}                 %%
             \end{table}}                     %%
\renewcommand{\title}[1]{\vbox{\center\LARGE{#1}}\vspace{5mm}}
\renewcommand{\author}[1]{\vbox{\center#1}\vspace{5mm}}
\newcommand{\address}[1]{\vbox{\center\em#1}}

\numberwithin{equation}{section}

\usepackage{hyperref}

\usepackage[numbers,sort&compress]{natbib}
\usepackage{hypernat}

\newcommand{\myref}[1]{}

\newcommand {\la} {\left \langle}
\newcommand {\ra} {\right \rangle}
\newcommand {\lb} {\left (}
\newcommand {\rb} {\right )}

\newcommand{\ket}[1]{\left |  #1 \right \rangle}

\newcommand {\CalO} {\mathcal O}

\newcommand {\CalN} {\mathcal N}

\newcommand {\CalM} {\mathcal M}

\newcommand {\BR}   {\mathbb R}

\newcommand {\BC}   {\mathbb C}
\newcommand {\BO}   {\mathbb O}

\newcommand {\CP}   {\mathbb C \mathbb P}

\newcommand {\ve}  {\varepsilon}
\newcommand {\ep}  {\epsilon}

\newcommand{\g}{\mathfrak{g}}

\renewcommand{\Re} {\mathrm{Re}}
\renewcommand{\Im} {\mathrm{Im}}

\newcommand {\p} {\partial}

\DeclareMathOperator{\tr} {tr}

\DeclareMathOperator{\Pexp} {Pexp}
\DeclareMathOperator{\ad} {ad}

\newcommand{\SU}{SU}
\newcommand{\U}{U}
\newcommand{\SO}{SO}
\newcommand{\Spin}{Spin}

\newcommand{\OSp}{OSp}

\newcommand{\PSU}{PSU}

\newcommand{\so}{\mathfrak {so}}
\newcommand{\osp}{\mathfrak {osp}}
\newcommand{\spn}{\mathfrak {sp}}
\newcommand{\su}{\mathfrak {su}}

\newcommand{\const}{\mathrm{const}}

\begin{document}

\unitlength = .8mm

\bibliographystyle{utphys}

\begin{titlepage}
\begin{center}
\hfill \\
\hfill \\
\vskip 1cm

\preprint{ITEP-TH-18/09}

\title{Localization of the four-dimensional $\CalN=4$ SYM to a two-sphere
and 1/8 BPS Wilson loops}

\renewcommand{\thefootnote}{\fnsymbol{footnote}}

\author{Vasily Pestun\footnotemark} 

\address{
%Center for the Fundamental Laws of Nature
%\\
Jefferson Physical Laboratory, Harvard University, Cambridge, MA 02138 USA}

%\email{$^a$pestun (at) physics.harvard.edu}

\footnotetext{On leave of absence from ITEP, 117218, Moscow, Russia}

\end{center}

\abstract{
We localize the   four-dimensional $\CalN=4$ super Yang-Mills theory on $S^4$ 
to the two-dimensional constrained Hitchin/Higgs-Yang-Mills (cHYM) 
theory on $S^2$.
We show that expectation values of certain 1/8 BPS supersymmetric Wilson loops on $S^2$ in the 4d $\CalN=4$ SYM is captured by
the 2d cHYM theory. We further argue that expectation values of Wilson 
loops in the cHYM theory agree with the prescription ``two-dimensional 
bosonic Yang-Mills excluding instanton contributions''. Hence, 
we support the recent conjecture by Drukker, Giombi, 
Ricci and Trancanelli on the 1/8 BPS Wilson loops on $S^2$ in the 4d $\CalN=4$ SYM. 
}

\vfill

\end{titlepage}

\renewcommand{\thefootnote}{\arabic{footnote}}
\setcounter{footnote}{0}

\eject

\tableofcontents

%%% Local Variables: 
%%% mode: latex
%%% TeX-master: "main"
%%% End: 

%\input{notestart}

\newcommand{\mSol}{\CalM_{2d}}
\newcommand{\mSolG}{\CalM'}

%\title{Wilson loops on $S^2$ subspace in four-dimensional $\CalN=4$ super Yang-Mills}%

%617 642 3873

\section{Introduction}

The dynamics of the $\CalN=4$ super Yang-Mills theory in four dimensions is probably 
the simplest among the four-dimensional gauge theories, but still this is a very rich and
 interesting theory from the theoretical perspective. 

Integrable structures discovered in the $\CalN=4$ SYM \cite{Minahan:2002ve}
or possible connection with the geometric Langlands program \cite{Kapustin:2006pk}
are just a few examples of interesting mathematics represented by this maximally 
supersymmetric gauge theory. It is believed that the theory has the exact dual description 
--- the type IIB ten-dimensional string theory in the $AdS_{5} \times S^{5}$ 
background \cite{Maldacena:1998re,Witten:1998qj,Gubser:1998bc}. 

The basic observables in any gauge theory, the Wilson loop observables, 
in the $\CalN=4$ theory can be generalized to preserve some amount of superconformal symmetry.
The simplest operator of this kind is a circular Wilson loop which  
couples to one of the six adjoint scalar fields of the $\CalN=4$ SYM.  Such operator
is called 1/2 BPS Wilson loop because it preserves one half of the 32 superconformal symmetries 
of the $\CalN=4$ SYM. In the beautiful work \cite{Erickson:2000af} further elaborated in 
\cite{Drukker:2000rr} it was conjectured that the expectation value of such operator 
can be computed in the Gaussian matrix model. In \cite{Pestun:2007rz} 
it was shown that this conjecture follows from localization of the path integral to 
the supersymmetric configurations. From the dual string theory point of view, in a suitable limit of large $N$ and 
large 't Hooft constant $\lambda = g_{YM}^2 N$, these Wilson loop observables are usually described 
by a string worldsheet with a boundary located at the loop \cite{Drukker:1999zq}. 
As was shown in the original paper \cite{Erickson:2000af}, the large $N$ and the large $\lambda$ limit
of the Gaussian matrix model nicely agrees with the solution to the minimal area problem 
in the dual string theory. 

In \cite{Zarembo:2002an,Dymarsky:2006ve,Drukker:2007qr,Drukker:2007dw,Drukker:2007yx}
other kinds of Wilson loops, which preserve various amount of supersymmetry, have been studied. 
In particular \cite{Drukker:2007qr,Drukker:2007yx,Drukker:2007dw} have constructed 1/16 BPS Wilson 
loops of arbitrary shape  on a three-sphere $S^3$ in the Euclidean space-time $\BR^4$
in the 4d $\CalN=4$ SYM. Restricting these Wilson loops to the equator $S^2 \subset S^3$ one gets 1/8 BPS
Wilson loops. In \cite{Drukker:2007qr,Drukker:2007yx,Drukker:2007dw} a bold conjecture has been proposed:
the expectation value of such Wilson loops is captured by the zero-instanton sector of the ordinary bosonic 
two-dimensional Yang-Mills living on the $S^2$. The coupling constant of the bosonic 2d YM 
is related to the coupling constant of the $\CalN=4$ SYM as $g_{2d}^2 = - g_{4d}^2 / (2 \pi r^2)$ where 
$r$ is the radius of the $S^2$. This conjecture was further supported at 
the order $\lambda^2$ in \cite{Young:2008ed,Bassetto:2008yf} for an expectation 
value of a single Wilson loop operator of arbitrary shape on $S^2$, but
in \cite{Young:2008ed} a discrepancy was found  at the order $\lambda^3$ for 
a connected correlator of two circular concentric Wilson loops on $S^2$.

Clearly, to support or to refine the conjecture one needs a framework which allows
to deduce this conjectural two-dimensional theory from the 4d  $\CalN=4$ SYM. 

In this paper we use localization argument to explain how the dynamics
of the Wilson loops on $S^2$ in the $d=4$ $\CalN=4$ SYM is captured
by a two-dimensional theory. 
Usually, the localization involves two steps (compare with e.g. \cite{Pestun:2007rz}): 
(i) finding out the configurations on which the theory localizes and evaluating
the physical action on these configurations, (ii) computing the determinant 
for the fluctuations of all fields in the normal directions
to the localization locus. We give the details of the step (i), leaving out the
step (ii) for future research. Compared with \cite{Pestun:2007rz}, where
the computation of the determinant was possible using the theory of indices
for transversally elliptic operators on compact manifold, in the present
case the complication is that the relevant operator in not transversally elliptic 
everywhere. However, this non-ellipticity is rather mild: the operator degenerates
at the codimension two submanifold of the space-time -- this is precisely 
the $S^2$ where the interesting Wilson loop operators are located. That gives a hope
that these  complications could be overpassed in a future.

In the localization we first use the circle action of the square of the
relevant supersymmetry generator to reduce the theory from
four-dimensions to three-dimensions. Next we study the supersymmetry equations
on the resulting 
three-dimensional manifold with a boundary $S^2$. The interesting Wilson loop
observables live at this two-dimensional boundary. The three-dimensional 
equations are quite complicated, but one can relate these equations
and the extended Bogomolny equations which appeared in \cite{Kapustin:2006pk}. 
We do not study singular solutions to these equations, but in principle, this
can be done, and it would correspond to the insertion of the 't Hooft operators
running over the circles linked with the two-sphere.  

Then we show that the moduli space of solutions to the supersymmetry equations is parametrized by the boundary data
and that the three-dimensional action on the supersymmetric solutions is captured by the boundary term. This boundary term is 
effectively the action of the two-dimensional theory living on the boundary.

The resulting
two-dimensional theory is the semi-topological Hitchin/Higgs-Yang-Mills theory (see
e.g. \cite{Moore:1997dj,Gerasimov:2006zt,Gerasimov:2007ap}).
We argue, though not totally rigorously, that the perturbative computation of the Wilson loop operators
in the HYM theory agrees with the perturbative computation in the usual 2d
bosonic YM, and that the unstable instantons in the HYM theory do not
contribute to the partition function  because of
extra fermionic zero modes. 

In other words, using the localization, we derive a Lagrangian formulation of 
the 2d theory which is supposed to capture Wilson loops on $S^2$ in the 4d $\CalN=4$ SYM, and we 
support the prescription ``the 0-instanton sector in the 2d bosonic Yang-Mills''
suggested in \cite{Drukker:2007qr,Drukker:2007yx,Drukker:2007dw}.

We have not found the determinant of quantum
fluctuations at the localization locus in this work, 
but there are good reasons to believe that this determinant 
in the $\CalN=4$ theory is trivial like it happened in \cite{Pestun:2007rz}.
In this case, and if one shows rigorously that the 2d HYM theory is equivalent 
to the ``zero-instanton sector'' of the 2d bosonic Yang-Mills for correlation 
function of Wilson loop operators, the conjecture of 
\cite{Drukker:2007qr,Drukker:2007yx,Drukker:2007dw} would be proved. It would be
in a nice agreement with several recent computations on 1/8 BPS Wilson loops made
in \cite{Giombi:2009ms,Bassetto:2009rt}. However, then we will have a puzzle 
how to reconcile this result with the explicit Feynman diagram computations at the order $\lambda^3$
for a connected correlator of two Wilson loops done at \cite{Young:2008ed} which were shown 
not to agree with the 2d YM conjecture. Perhaps, there are involved subtle issues related to the regularization 
of the conformal supersymmetric gauge theory and/or anomalies which require further studies in either approach
to the problem. 

Another scenario is that the HYM theory is corrected by the one-loop determinant.
However,  this 
correction must be constrained by the following results: (i) it could show up only at the order of $\lambda^3$ for the correlators of  Wilson loops 
of arbitrary shape but not at smaller order \cite{Young:2008ed,Bassetto:2008yf}, 
(ii) in the large $\lambda$ and the large $N$ limit this correction has to vanish because the dual 
string computation agrees with the matrix model computation for the connected correlator which 
follows from the conjecture \cite{Giombi:2009ms}, (iii) this correction 
must not contribute to the expectation value of the Wilson loop operator on the equator on $S^2$ 
which was proved to be computed by the Hermitian matrix model \cite{Pestun:2007rz} (and 
this Hermitian matrix model is implied by the conjecture).
Future research is needed to resolve these interesting issues.

In section \ref{se:geom-setup} we describe the geometry of the
Wilson loops which we study together with the relevant supersymmetries 
and also set up various notations and conventions. A reader well familiar with constructions in \cite{Drukker:2007qr,Drukker:2007yx,Drukker:2007dw} might wish to skip straight to the section \ref{se:localization} where we describe the actual localization computation. In section \ref{se:two-dimensional} we analyze the resulting two-dimensional theory.

\subsection*{Acknowledgements}
 A part of this work has been done while completing Ph.D. thesis 
of the author at Princeton University under advisement of Edward Witten. 
 The author is grateful to  Anatoly Dymarsky, Simone Giombi, Nikita Nekrasov, Samson Satashvili and Edward Witten for interesting discussions on this project.
This work has been partially supported by Jacobus Honorific Fellowship
at Princeton University, a Junior Fellowship at Harvard Society of Fellows, and grants
NSh-3035.2008.2 and RFBR 07-02-00645.

\section{The  conventions and geometry\label{se:geom-setup}}

Let $X_i$, $i =1 \dots 5$, be coordinates in $\BR^5$ into which the $S^4$ is
embedded as the hypersurface $\sum X_i^2 = r^2$.
By $x_i$, $i =  1\dots 4$, we denote the standard
coordinates on the stereographic projection from $S^4$ to $\BR^4$ which maps
 the North pole $N$ of the $S^4$ with  coordinates $\vec X=(0,0,0,0,r)$ to the origin  $\vec x = 0$ of the $\BR^4$:
\begin{equation}
\begin{aligned}
  \label{eq:X-x-coordinates-relations}
  X_i & = \frac {x_i}{ 1 + \frac {x^2}{4 r^2}},\quad i = 1\dots 4  \\
  X_5 & = r \frac { 1 - \frac {x^2}{4 r^2}} {1 + \frac {x^2}{4 r^2}}.
\end{aligned}
\end{equation}

We define the three-sphere $S^3 \subset S^4$ by the equation $X_5 =
0$. Equivalently, in the $x_i$ coordinates on $\BR^4$, this three-sphere is defined
by the equation $x^2 = 4r^2$. Next, we define the two-sphere $S^2 \subset
S^3$ by the additional equation $X_1 = 0$. In the $x_i$ coordinates, the $S^2$
is described by the equations $\{x_1 =0, x_2^2 + x_3^2 + x_4^2 = 4 r^2 \}$.
We denote this $S^2$ as $\Sigma$.

We call the point $P$ with $\vec X(P) = (0,r,0,0,0)$ the North pole of $\Sigma$.
(The points $P$ and $N$ are different points). In $x^i$ coordinates, $\vec x(P) = (0, 2r, 0,0)$.
By $y_i$, $i = 1 \dots 4$, we denote the standard coordinates on the stereographic
projection from $S^4$ to $\BR^4$ which maps the point $P$ to the origin of the $\BR^4$:
\begin{equation}
  \label{eq:X-y-coordinates-relation}
  \begin{aligned}
    X_i& = \frac { y_i }{ 1 + \frac {y^2} { 4r^2}}, \quad i = 1,3,4 \\
    X_5& = \frac {-y_2} { 1 + \frac {y^2} { 4r^2}} \\
    X_2& = r \frac { 1 - \frac {y^2}{4 r^2}} { 1 + \frac {y^2}{4r^2}}.
  \end{aligned}
\end{equation}

The $\SO(5)$ isometry group of $S^4$ can be broken to $\SO(2)_S \times \SO(3)_S$
where the $\SO(2)_S$ acts on $(X_1,X_5)$ and  the $\SO(3)_S$ acts on $(X_2, X_3, X_4)$.\footnote{We shall use the subscript ''S'' to denote subgroups of
the space-time symmetries, and the subscript ''R'' do denote subgroups of the
$R$-symmetry. We also remark that the $\SO(3)_S$ subgroup of the $\SO(4)$
isometry group of $\BR^4$ is not the left $\SU(2)_L$ subgroup in the decomposition
$\SO(4) = \SU(2)_L \times SU(2)_R$, but rather a diagonal embedding.}
The two-sphere $\Sigma$ is the fixed point set of the $\SO(2)_S$.
Sometimes it is convenient to use the  $\SO(2)_S \times \SO(3)_S$ spherical coordinates on $S^4$; we represent the $S^4$ as a warped $S^2 \times S^1$  fibration over an interval $I$. Let  $\theta \in [0,\pi/2]$ be the coordinate on $I$. 
We also use notation $\xi = \pi/2 - \theta$. Let $\tau \in [0, 2 \pi)$   be the standard coordinate on $S^1$ fibers and let $d \Omega_2^2$ 
be the standard unit metric on the $S^2$ fibers. Then the metric on $S^4$ of radius $r$ is
form 
\begin{equation}
  \label{eq:metric-polar-coordinates-S4}
  ds^2 = r^2(d \theta^2  + \sin^2 \theta \, d\tau^2 + \cos^2 \theta \, d\Omega_2^2)
\end{equation}
At $\theta =0$ the $S^1$  shrinks
to zero and the $S^2$ is of maximal size, while at $\theta = \pi /2$ the $S^2$ shrinks to
zero and the $S^1$ is of maximal size.

\subsection{1/8 BPS Wilson loop operators}

\myref{2008-05-06 p6} \myref{2008-05-01 p6} Following \cite{Drukker:2007qr,Drukker:2007yx,Drukker:2007dw} we
consider  the Wilson loops located on the $S^3$ 
of the following form 
\begin{equation}
  \label{eq:Wilson-loops-S3}
  W_R (C) =  \tr _R \Pexp \oint \left(A_{\mu} + i \sigma_{\mu \nu}^{A} \frac
  {x^{\nu}}{2r} \Phi_A \right) dx^{\mu},
\end{equation}
specifically restricting our attention to the Wilson loops located
on the equator $\Sigma = S^2 \subset S^3$.
The definition (\ref{eq:Wilson-loops-S3}) is given
in the $\BR^4$ stereographic coordinates $x_i$ 
(\ref{eq:X-x-coordinates-relations}).\footnote{Recall that in our conventions
the equation of the $S^3$ is $\sum_{i=1}^{4} x_i^2 = 4 r^2$.}

The definition of such Wilson loops and the condition for supersymmetry
was found in \cite{Drukker:2007dw,Drukker:2007yx,Drukker:2007qr}.
The $\Phi_A$ denotes  three of six scalar fields of the $\CalN=4$ super Yang-Mills theory. 
In our conventions the index $A$ runs over $6,7,8$. The $\mu,\nu$ are the space-times indices running over $1,\dots 4$. 
The $\sigma^{A}_{\mu \nu}$ are the 't Hooft symbols:
three $4\times 4$ anti-self-dual matrices satisfying $\su(2)$ commutation
relations. Explicitly we choose
\begin{equation}
  \label{eq:my-sigma-definition-for-Wilson-loops}
  \begin{aligned}
    \sigma^{i+4}_{1 i} &= 1  &    \sigma^{i+4}_{jk}  &= - \ep_{ijk} \quad \text{for} \quad i = 2,3,4,
  \end{aligned}
\end{equation}
where $\ep_{ijk}$ is the standard
antisymmetric symbol with $\ep_{234} = 1$.
The $\SO(6)$ R-symmetry group is broken to $\SO(3)_A \times SO(3)_B$. Our
conventions are that the $\SO(3)_A$ acts on the three scalars
$\Phi_6,\Phi_7, \Phi_8$ which couple to the Wilson loop
(\ref{eq:Wilson-loops-S3}). The
$\SO(3)_B$ acts on the remaining scalars $\Phi_5, \Phi_9, \Phi_0$.
The Wilson loop (\ref{eq:Wilson-loops-S3}) is explicitly invariant under the
$\SO(3)_B$ symmetry, because the scalar fields $\Phi_5,\Phi_9, \Phi_0$ do not
appear in (\ref{eq:Wilson-loops-S3}).
In the case when the Wilson loop (\ref{eq:Wilson-loops-S3}) is restricted to the
two-sphere $S^2$ by the constraint $x_1 = 0$, it is also invariant under the
diagonal $\SO(3)$ subgroup of the $\SO(3)_S
\times \SO(3)_A$, i.e. under the simultaneous rotation of the coordinates $x_i$ and
the scalars $\Phi_{i+4}$, $i=2,3,4$.

The supersymmetries which are preserved by the Wilson loop
(\ref{eq:Wilson-loops-S3}) were found in \cite{Drukker:2007qr,Drukker:2007yx,Drukker:2007dw}. \myref{2008-05-06 p7} To set all notations and
conventions we repeat the derivation here.

\subsection{Superconformal symmetries and conformal Killing spinors}

The conformal Killing spinor on $\BR^4$ is parameterized by two constant spinors
which we call $\hat \ve_{s}$ and $\hat \ve_{c}$, where $\hat \ve_{s}$ generates the usual
Poincare supersymmetries, and $\hat \ve_{c}$ generates the special superconformal
symmetries
\begin{equation}
  \label{eq:conformal-Killing-spinor}
  \ve(x) = \hat \ve_{s} + x^{\rho} \Gamma_{\rho}\hat \ve_{c}.
\end{equation}

The variation of the bosonic fields of the theory is
\begin{equation}
  \label{eq:variation-of-bosonic-fields}
  \delta A_{M} = \psi \Gamma_{M} \ve,
\end{equation}
where $A_M$, $M=0,\dots, 9$ is a collective notation for the gauge fields $A_{\mu}$, $\mu = 1,\dots,4$,
and the scalar fields $\Phi_{A}$, $A =5, \dots, 9,0$. The $\psi$ denotes sixteen component fermionic fields of the $\CalN=4$ theory 
written in the $d=10 $ $\CalN=1$ SYM notations.  The $\Gamma_M$, $M=0,\dots,9$, are $16\times 16$ gamma-matrices which
act on the chiral spin representation $S^{+}$ of $\Spin(10)$. The spinors $\psi, \ve, \hat \ve_{s}$ are in the $S^{+}$
while $\hat \ve_{c}$ is in the $S^{-}$. 
The variation of a generic Wilson loop (\ref{eq:Wilson-loops-S3}) vanishes if
and only if $\ve$
satisfies
\begin{equation}
  \label{eq:variation-of-Wilson-loop}
  (\Gamma_{\mu} + i \Gamma_A \sigma^{A}_{\mu \nu} \frac {x^{\nu}}{2 r}) (\hat
  \ve_{s} + x^{\rho} \Gamma_{\rho} \hat \ve_{c}) \dot x^{\mu}= 0
\end{equation}
for any point $x \in S^{3}$ and the tangent vector $\dot x_{\mu}$ which is constrained by $\dot x_{\mu} x_{\mu} = 0$. The terms linear in $x$ give the equation
\begin{equation}
 x^{\mu} \dot x^{\rho}( \Gamma_{\mu} \Gamma_{\rho} \hat \ve_{c} + i \Gamma_{A} \sigma^{A}_{\mu \rho}
  \frac {\hat \ve_s} { 2 r}) =0.
\end{equation}
Since the vectors $x^{\mu}$ and $\dot x_{\mu}$ are constrained only by $x^{\mu}
\dot x_{\mu} = 0$, we get
\begin{equation}
  \label{eq:constraint-1-on-ve}
  \Gamma_{\mu \rho} \hat \ve_{c} + i \Gamma^{A} \sigma^{A}_{\mu \rho} \frac
  {\hat \ve_{s}} {2r} = 0.
\end{equation}
The constant and quadratic in $x$ terms give the equation
\begin{equation}
 \dot x^{\mu} ( \Gamma_{\mu} \hat \ve_{s} + i \Gamma_{A \lambda} \frac {\sigma^{A}_{\mu \nu}}{2r} x^{\nu}
  x^{\lambda} \hat \ve_{c}) = 0.
\end{equation}
Multiplying by non-degenerate matrix $x^{\rho} \Gamma_{\rho}$ we get
\begin{equation}
  \dot x^{\mu} x^{\rho} ( \Gamma_{\rho \mu} \hat \ve_{s}  + i \Gamma_{\rho}
  \Gamma_{A \lambda} \frac {\sigma^A_{\mu \nu}}{2r}  x^{\nu}
  x^{\lambda} \hat \ve_{c}) = 0.
\end{equation}
Using $x^{\mu} x_{\mu} = 4 r^2$ and $\dot x^{\mu} x_{\mu} =0 $ we get
\begin{equation}
 \label{eq:constraint-2-on-ve}
 \Gamma_{\mu \rho} \hat \ve_{s} + i \Gamma_{A} \sigma^{A}_{\mu \rho} (2r) \hat
 \ve_{c} = 0.
\end{equation}
The equation (\ref{eq:constraint-2-on-ve}) is actually equivalent to
(\ref{eq:constraint-1-on-ve}) and to
\begin{equation}
  \label{eq:combined-equation-on-eps}
  2r \hat \ve_{c} = i \sigma^{A}_{\mu \rho} \Gamma_{A \mu \rho} \hat \ve_{s}.
\end{equation}
If Wilson loop is restricted to $S^2$, then (\ref{eq:combined-equation-on-eps})
amounts to three maximally orthogonal projections in the spinor representation space
$S^{+} \oplus S^{-}$. Each projection operator reduces the dimension of the
space of solutions by half. Starting from the dimension 32 of $S^{+} \oplus
S^{-}$ we get $32/2^3 = 4$-dimensional space of solutions for $(\hat \ve_{s},
\hat \ve_{c})$. For generic Wilson loops on $S^3$ the dimension of the space of
solutions is further reduced by two, so there are only $2$ supersymmetries left.

For explicit computation we use the following $16 \times 16$ gamma-matrices representing
Clifford algebra on $S^{+}$:
\begin{equation}
  \label{eq: gamma-matrices}
  \begin{aligned}
    &\Gamma_{M} =
      \begin{pmatrix}
        0 & E_{M}^{T} \\
        E_{M} & 0 \\
      \end{pmatrix}
    , \quad  M=2 \dots 9 \\
    &\Gamma_1 =
      \begin{pmatrix}
        1_{8\times8} & 0 \\
        0 & -1_{8\times 8} \\
      \end{pmatrix}
    , \\
    &\Gamma_0 =
      \begin{pmatrix}
        i 1_{8\times8} & 0 \\
        0 & i 1_{8\times 8} \\
      \end{pmatrix}
    ,
  \end{aligned}
\end{equation}
Here $E_{M}$, $M=2\dots 8$, are  $8\times8$ matrices representing left
multiplication of the octonions and $E_{9}=1_{8\times8}$. (Let $e_{i}$, $i=2,\dots, 9$, be the generators of the octonion algebra $\BO$. We chose $e_9$ to be
identity.  Let $c^k_{ij}$ be the structure
constants of the left multiplication $e_i  \cdot e_j = c^{k}_{ij} e_k$. Then
$(E_i)^k_{j} = c^{k}_{ij}$. The multiplication table is defined by specifying
the set of cyclic triples $(ijk)$ such that $e_i e_j = e_k$. We define the cyclic triples
to be $(234),(256),(357),(458),(836),(647),(728)$.)

Explicitly, the four linearly independent solutions of
(\ref{eq:combined-equation-on-eps}), i.e. supersymmetries of Wilson loops on the
$S^2$ are the following
\newcommand{\veunit}{\ket{1}}
\begin{equation}
\begin{aligned}
  \label{eq:explicit-solutions-for-epsilon-S2-at-N}
  \hat \ve^s_1 =
  \begin{pmatrix}
    1 \\  0 \\ -1 \\  0
  \end{pmatrix} \otimes \veunit \quad
  \hat \ve^s_2 =
  \begin{pmatrix}
    0 \\ 1 \\ 0 \\ 1
  \end{pmatrix} \otimes \veunit \quad
  \hat \ve^s_{\bar 1} =
  \begin{pmatrix}
    1 \\ 0 \\ 1 \\ 0
  \end{pmatrix} \otimes \veunit \quad
  \hat \ve^s_{\bar 2} =
  \begin{pmatrix}
     0 \\ 1 \\ 0 \\ -1
  \end{pmatrix} \otimes \veunit \quad  \\
  \hat \ve^c_1 = \frac {1} {2r}
  \begin{pmatrix}
    0 \\  i \\ 0 \\  i
  \end{pmatrix} \otimes \veunit \quad
  \hat \ve^c_2 = \frac {1} {2r}
  \begin{pmatrix}
    -i \\ 0 \\ i \\ 0
  \end{pmatrix} \otimes \veunit \quad
  \hat \ve^c_{\bar 1} = \frac {1} {2r}
  \begin{pmatrix}
    0 \\ -i \\ 0 \\ i
  \end{pmatrix} \otimes \veunit \quad
  \hat \ve^c_{\bar 2} =  \frac {1} {2r}
  \begin{pmatrix}
     i \\ 0 \\ i \\ 0
  \end{pmatrix} \otimes \veunit \quad  \\
\end{aligned}.
\end{equation}
Sixteen components of the spinors are written in the $4
\times 4$ block notations, where
\begin{equation}
  \veunit =
  \begin{pmatrix}
    1 \\ 0 \\ 0 \\ 0
  \end{pmatrix}.
\end{equation}
In  more generic case of Wilson loops on $S^3$, one gets only the two-dimensional
space of solutions \cite{Drukker:2007qr}, which is spanned by $\ve_{1}, \ve_{2}$,
but not by $\ve_{\bar 1}, \ve_{\bar 2}$.\footnote{
We use indices $1,2$ and $\bar 1, \bar 2$ only to enumerate the basis elements of the solutions to (\ref{eq:combined-equation-on-eps}), but it
 is not supposed that $\ve_{\bar 1}$ or $\ve_{\bar 2}$ is the complex conjugate to $\ve_{1}$ or
$\ve_{2}$.}

\subsection{Anticommutation relations }

 Let $Q_{1}, Q_{2}, Q_{\bar 1}, Q_{\bar 2}$ be the four conformal supersymmetries
generated by conformal Killing spinors (\ref{eq:conformal-Killing-spinor}) with
$\hat \ve_{s}, \hat \ve_{c}$ given by
(\ref{eq:explicit-solutions-for-epsilon-S2-at-N}).
Let $R_{AB}$ be the matrices in the fundamental representation of the $SO(6)$
R-symmetry generators. On scalar fields the generators $R_{AB}$ act as
\begin{equation}
  \label{eq:SO-generators-on-scalar-fields}
  (\delta_{R_{AB}} \Phi)_{A} = R_{AB} \Phi_B.
\end{equation}
The fermionic symmetries anticommute according to (\ref{eq:delta2onAPhi}),(\ref{eq:delta4fermions})
as
\begin{equation}
  \delta_{\ve}^2 \Phi_A = 2 (\tilde \ve \Gamma_{AB} \ve) \Phi_B,
\end{equation}
hence the R-symmetry part of the anticommutators is
\begin{equation}
  Q_{\{\alpha} Q_{\beta \}}  = 2 (\tilde \ve_{\{\alpha} \Gamma_{AB}
  \ve_{\beta\}}) R_{AB}.
\end{equation}
For space-time rotations we have similar equation except for the sign. Let us
consider a fixed point of the space-time rotation. Then, assuming that the
$\SO(4)_S$ generators $R_{\mu \nu}$ act on tangent space $\BR^4$ in the same way
as the $\SO(6)_R$ generators $R_{AB}$ act on the scalar target space $\BR^{6}$,
we get the space-time symmetry part of the anticommutators
\begin{equation}
  Q_{\{\alpha} Q_{\beta \}}  = -2 (\tilde \ve_{\{\alpha} \Gamma_{\mu \nu}
  \ve_{\beta\}}) R_{\mu \nu},
\end{equation}
where $\ve$ and $\tilde \ve$ are taken at the fixed point set of the space-time rotation.
To summarize,
\begin{equation}
  Q_{\{\alpha} Q_{\beta \}}  = 2 (\tilde \ve_{\{\alpha} \Gamma_{AB}
  \ve_{\beta\}}) R_{AB} -2 (\tilde \ve_{\{\alpha} \Gamma_{\mu \nu}
  \ve_{\beta\}}) R_{\mu \nu}.
\end{equation}
At a fixed point of space-time rotation, the $SO(4)_S \times \SO(6)_R$
generators act on spinors in the $S^{+}$ representation of $\SO(10)$ as
\begin{equation}
  \label{eq:R-MN-action-on-spinors}
  \delta_{R_{MN}} \Psi = \frac 1 4 R_{MN} \Gamma^{MN} \Psi.
\end{equation}

Then there are the following
anticommutation relations \myref{2008-05-02 p13} \myref{2008-05-06 p1}
\myref{2008-05-03 p5}
\begin{equation}
\begin{aligned}
  \{Q_1, Q_1\} & = \frac 2 r  R_{05} - \frac 2 r  i R_{59} &  \{Q_{\bar 1}, Q_{\bar 1} \} & = \frac 2 r
  R_{05} + \frac 2 r  i R_{59} \\
  \{Q_2, Q_2\} & = -\frac 2 r  R_{05} - \frac 2 r  i R_{59} & \{Q_{\bar 2}, Q_{\bar 2} \} & = -\frac 2 r
  R_{05} + \frac 2 r  i R_{59}  \\
  \{Q_1, Q_2 \} &= \frac 2 r  R_{09} & \{Q_{\bar 1},Q_{\bar 2} \} &= - \frac 2 r  R_{09} \\
  \{Q_1, Q_{\bar 1} \} & = -\frac 2 r  R_{12} & \{Q_{1}, Q_{\bar 2}\} &= 0 \\
  \{Q_2, Q_{\bar 1} \} & = 0  & \{Q_{2}, Q_{\bar 2}\}  & = -\frac 2 r  R_{12}
\end{aligned}.
\end{equation}

These anticommutation relations can be packed into
\begin{equation}
  \label{eq:anti-commutators-SU2-relations}
  \begin{aligned}
  \{Q_{\alpha}, Q_{\beta}\} &= \frac {2} {r} (C \sigma^{I})_{\alpha \beta} R_{I} \\
  \{Q_{\bar \alpha}, Q_{\bar \beta}\} &= \frac {2}{r} (\overline{C \sigma^{I}
  })_{\bar \alpha \bar \beta} R_{I} \\
  \{Q_{\alpha}, Q_{\bar \beta}\} &= \frac 2 r \delta_{\alpha \bar \beta} R_0,
  \end{aligned}
\end{equation}
where $\sigma^{I}$, $I=1,2,3$, are the Pauli matrices
\begin{equation}
  \label{eq:Pauli-matrices}
  \sigma^1 =
  \begin{pmatrix}
    0 & 1 \\
    1 & 0
  \end{pmatrix} \quad
  \sigma^2 =
  \begin{pmatrix}
    0 & -i \\
    i & 0
  \end{pmatrix} \quad
  \sigma^3 =
  \begin{pmatrix}
    1 & 0 \\
    0 & -1
  \end{pmatrix}.
\end{equation}
The $C$ denotes ``the charge conjugation'' matrix, $C = i \sigma_2$,
the triplet of the $\SO(3)_B$ generators is denoted by $R_I$ such that
$(R_1,R_2,R_3) := (R_{05}, -R_{59}, -R_{09})$, and the $\SO(2)_S$ generator is
called $R_0 := -R_{12}$. \myref{2008-05-02 p19} 
%[$R_0 = -U$ there]

The fermionic generators $Q_{\alpha}$ and $Q_{\bar \alpha}$ transform naturally
in the representation $\bf{2}$ and $\bf{\bar 2}$ of the $\SO(3)_B \simeq SU(2)_B$, while
$\SO(2)_S$ mixes them
\myref{2008-05-01 p24} \myref{2008-05-02 p18} \myref{2008-05-02 p20}
\begin{equation}
\label{eq: R-Q-commutation-relations-SU1-2}
 \begin{aligned}[]
 [R_I Q_{\alpha}] &= -\frac 1 2 i \sigma^I_{\alpha \beta} Q_{\beta}  &
 [ R_0 Q_{\alpha}] & = - \frac 1 2 i C_{\alpha \bar \beta} Q_{\bar \beta} \\
 [R_I Q_{\bar \alpha}] &= \frac 1 2 i \bar \sigma^{I}_{\bar \alpha \bar \beta} Q_{\bar \beta} &
 [ R_0 Q_{\bar \alpha}] &= \frac 1 2 i C_{\bar \alpha \beta} Q_{\beta}.
 \end{aligned}
\end{equation}

The  relations (\ref{eq:anti-commutators-SU2-relations}) and (\ref{eq:
  R-Q-commutation-relations-SU1-2}) are the
commutation relations of the Lie algebra $\su(1|2)$ of the $SU(1|2)$ subgroup of
the superconformal group \cite{Drukker:2007qr}.\myref{2008-05-03 p8} The bosonic part of $\su(1|2)$ is $\so(2)_S \times
\so(3)_B$, spanned by $R_0, R_I$, the fermionic part is
four-dimensional, spanned by $Q_{\alpha}, Q_{\bar \alpha}$.

If we take a linear combination of the fermionic generators with
complex coefficients $\ve^{\alpha},\ve^{\bar \alpha}$
\begin{equation}
  \label{eq:Q-definition}
  Q = \ve^{\alpha}Q_{\alpha} + \ve^{\bar \alpha} Q_{\bar \alpha},
\end{equation}
we will find that $Q$ squares to a real generator of the $\SO(3)_B \times
\SO(2)_S$ if $\ve^{\bar \alpha}$ is actually the complex conjugate to
$\ve^{\alpha}$. Such $Q$ will be called Hermitian
and will be used in the following for the localization
computation. We shall also notice that if $Q$ is Hermitian, i.e. if $\ve^{\bar \alpha}$ is complex conjugate to $\ve^{\alpha}$,
then the norm of the $\SO(2)_S$ generator and
$\SO(3)_B$ generator in $Q^2$ is proportional to the norm of $\ve$. Hence,
a non-zero Hermitian $Q$ always squares to a non-zero rotation generator in both
$\SO(2)_S$ and $\SO(3)_B$. \myref{2008-05-02 p23}

\subsubsection{The localization operator $Q$}

For explicit localization computations we take $Q$ to be
\begin{equation}
  \label{eq:generator-Q}
Q_{\ve} = \frac 1 2 (Q_{1} + Q_{\bar 1}).
\end{equation}
It corresponds to the conformal Killing spinor generated by 
\begin{equation}
  \label{eq:epsilon-s-epsilon-c-for-Q}
 \hat \ve_{s} =
 \begin{pmatrix}
   1 \\ 0 \\ 0 \\ 0
 \end{pmatrix} \otimes \veunit \quad
 \hat \ve_{c} = \frac {1} {2r}
 \begin{pmatrix}
   0 \\ 0 \\ 0 \\ i
 \end{pmatrix} \otimes \veunit.
\end{equation}
By (\ref{eq:anti-commutators-SU2-relations}) we have \myref{2008-05-06 p1}
\begin{equation}
  \label{eq:Q-squared}
  Q^2 = \frac {1} {r} (R_{05} - R_{12}).
\end{equation}
Clearly, since $[Q^2,Q] = 0$ we have
\begin{equation}
  \label{eq:R-on-Q-commmutation-vanishes}
  [R_{05} - R_{12}, Q] = 0  \quad \Rightarrow \quad (\Gamma_{05} - \Gamma_{12})
  \ve_{P} = 0.
\end{equation}
The last equality is written for the conformal Killing spinor $\ve$ associated
with $Q$ at the point $P$ in coordinate patch $y^{\mu}$
(\ref{eq:X-y-coordinates-relation}). The rotation of $(y_1,y_2)$ plane
corresponds in the global coordinates to the rotation of $(X_5, X_1)$ plane, or
the vector field $\frac{\p} {\p \tau}$ in the polar coordinates
(\ref{eq:metric-polar-coordinates-S4}). Geometrically, the equation
(\ref{eq:R-on-Q-commmutation-vanishes}) means that the conformal Killing spinor
$\ve$ is invariant under simultaneous rotation of the $(X_5, X_1)$ plane and
$(\Phi_5, \Phi_0)$ plane.

From the condition (\ref{eq:combined-equation-on-eps}) on $\ve$ and
(\ref{eq:my-sigma-definition-for-Wilson-loops}) it follows that $\ve$ is also
invariant under the diagonal rotations in the $\SO(3)_S \times \SO(3)_A$. Indeed,
from (\ref{eq:combined-equation-on-eps}) one gets
\begin{equation}
  \label{eq:invariance-of-epsilon-under-diagonal-SO3S-SO3A}
  \Gamma_{j+4} \Gamma_{ki} \hat \ve_{s} = \Gamma_{i+4} \Gamma_{jk} \hat \ve_{s}
\end{equation}
for pairwise distinct indices $i,j,k$ running over $2,3,4$. Multiplying by $\Gamma_{j+4}
\Gamma_{jk}$ both sides of this equation we get
\begin{equation}
  \label{eq:invariance-of-epsilon-under-diagonala-SO3S-SO3A-shown}
  \Gamma_{ji} \hat \ve_{s} = -\Gamma_{j+4, i+4} \hat \ve_{s},
\end{equation}
which shows that $\ve$ is invariant under simultaneous $\SO(3)_S$ rotation of
$(X_2,X_3,X_4)$ and the corresponding
$\SO(3)_A$ rotation of $(X_6,X_7,X_8)$ under the isomorphism $\BR^3 \to \BR^3: X_{i} \mapsto X_{i+4}$.

\subsubsection{Remark on 1/16 BPS Wilson loops on $S^3$}

We shall remark that a generic supersymmetric Wilson loop on the $S^3$ is
invariant only under the $\OSp(1|2, \BC)$ subgroup of the
complexified superconformal group $\PSU(2,2|4,\BC)$, see \cite{Drukker:2007qr}.
The fermionic part of $\OSp(1|2, \BC)$ is spanned by $Q_{\alpha}$, i.e. by half
of generators of $\SU(1|2, \BC)$. The bosonic part of $\osp(1|2,\BC)$ is
$\spn(2,\BC) \simeq \su(2,\BC)$ spanned by $R_{I}$. The commutation relations
are represented by the first equation in (\ref{eq:anti-commutators-SU2-relations}) and
in (\ref{eq: R-Q-commutation-relations-SU1-2}).
However, there is no real structure on $\OSp(1|2,\BC)$ such that the
real version of $\OSp(1|2,\BC)$ could be embedded into the compact 
unitary supergroup $\SU(1|2,\BR)$.\footnote{
If we use signature for $(5,9,0)$ directions $(+,+,-)$, then, since in this case
gamma-matrices can be chosen real, we can get a real structure on $\OSp(1|2,
\BR)$ by taking all generators to be real. However, in this case, $Q^2$ is
always light-like generator of the bosonic part of $\SO(2,1) \simeq SL(2,\BR)$ \myref{2008-05-03 p1}.}

 So there exists no fermionic element $Q$ in $\OSp(1|2,\BC)$ such that $Q^2$ generates a unitary
transformation.\myref{2008-05-01 p20} Since the localization
method, which we are using in this work, requires the global
transformation generated by $Q^2$ to be unitary, we cannot generalize
our localization computation to the $\OSp(1|2,\BC)$ case,
and, hence, we cannot treat generic Wilson loops on $S^3$ in the same way
as generic Wilson loops on the $S^2 \subset S^3$. So we
restrict the detailed study to the case of Wilson loops on $S^2 \subset S^3$.

\subsubsection{Remark on $1/4$ BPS circular Wilson loops}
As discussed above, a Wilson loop  (\ref{eq:Wilson-loops-S3})  of an arbitrary shape on $\Sigma=S^2$ preserves $4$
out of $32$ superconformal symmetries, so it can be called $4/32 = 1/8$ BPS 
Wilson loop, but a circular Wilson loop on $S^2$ preserves $8$ supersymmetries ($1/4$ BPS) \cite{Drukker:2007qr,Drukker:2007yx,Drukker:2007dw,Drukker:2006zk,Drukker:2006ga},
and the circular Wilson loop of maximal size preserves $16$ supersymmetries ($1/2$ BPS).
The Wilson loop on the equator of $S^2$ is the most
familiar maximally supersymmetric superconformal 
Wilson loop, the study of which was initiated in 
\cite{Erickson:2000af,Drukker:2000rr} and many consequent papers. 
There it was conjectured
that expectation value of such operator can be computed in a Gaussian matrix
model. In \cite{Drukker:2000rr} an argument was given that the field theory
localizes to matrix model, however that argument does not show that the matrix
model is Gaussian. In \cite{Pestun:2007rz} it was shown how 
to get the Gaussian matrix model from the localization computation. 

In \cite{Drukker:2006ga} it was conjectured that $1/4$ BPS circular Wilson loops
 can also be computed using the Gaussian matrix model but with a rescaled
coupling constant. Such $1/4$ BPS circular Wilson loops can be considered as an
intermediate case between maximally supersymmetric $1/2$ BPS Wilson loops and $1/8$
BPS Wilson loops of an arbitrary shape on $S^2$.

One may ask whether it is possible to localize the $\CalN=4$ SYM field theory for $1/4$ BPS
circular Wilson loops straight to the Gaussian matrix model?
We shall note that a new localization computation, different from localization
computation for a generic Wilson loop on $S^2$, might be possible only for a single circular
$1/4$ BPS Wilson on $S^2$. But if we take two $1/4$ BPS loops located
at two distinct latitudes $\beta_1$ and $\beta_2$ on $S^2$, then each Wilson loop
preserves eight supesymmetries, but only four supersymmetries are preserved
by both loops simultaneously. These four common supersymmetries are actually the
same as for a generic $1/8$ BPS Wilson loop on $S^2$. Hence, if we want to
compute the connected correlator of two latitudes on $S^2$, we are
back to the case of generic $1/8$ BPS loops on $S^2$, where the four-dimensional theory localizes
to a certain two-dimensional theory on $S^2$. So to compute correlator of two $1/4$ BPS circular 
Wilson loops we cannot localize the field theory straight to the two-matrix model~\cite{Giombi:2009ms}, 
but we have to deal with an intermediate two-dimensional field theory on $\Sigma$.

\subsection{Summary}
We study supersymmetric Wilson loops on $\Sigma=S^2 \subset S^3$ given by
 (\ref{eq:Wilson-loops-S3}). These Wilson loops are invariant under the
 $\SU(1|2)$ subgroup of the superconformal group, where $\U(1) = \SO(2)_S$
rotates $(X_1,X_5)$ plane, and $\SU(2) = \SU(2)_B$ rotates $(\Phi_5,\Phi_9,\Phi_0)$.
The Wilson loops are also invariant under the diagonal of $\SO(3)_S \times
\SO(3)_A$, where $\SO(3)_S$ acts on $(X_2,X_3,X_4)$ and $\SO(3)_A$ acts on
$(\Phi_6,\Phi_7,\Phi_8)$, i.e. on scalar fields appearing in the definition of Wilson loop.

We choose Hermitian generator $Q$, generated by the conformal Killing spinor
$\ve$, as in (\ref{eq:generator-Q}). The spinor $\ve$ is invariant under the
diagonal subgroup  of $\SO(3)_S
\times \SO(3)_A$ by
(\ref{eq:invariance-of-epsilon-under-diagonala-SO3S-SO3A-shown}) and the
diagonal subgroup of $\SO(2)_S
\times \SO(2)_B$ by (\ref{eq:R-on-Q-commmutation-vanishes}), where the $\SO(2)_B \subset
\SO(3)_B$ acts on $(\Phi_5, \Phi_0)$-plane. \myref{2008-05-04 p1} \myref{2008-05-06 p5}

\section{Localization\label{se:localization}}
\subsection{Introduction}

We want to show that the expectation value of the Wilson loops
(\ref{eq:Wilson-loops-S3}) on $\Sigma = S^2$ in four-dimensional $\CalN=4$
Yang-Mills can be computed in a certain two-dimensional theory localized to
$\Sigma$. The fermionic symmetry $Q$
(\ref{eq:generator-Q}) is BRST-like generator of equivariantly cohomological field theory, thanks to
the fact that $Q$ squares to global unitary transformation and gauge
transformation. This claim is true off-shell after adding to the theory the necessary
auxiliary fields. The operator $Q^2$ is the off-shell symmetry of the action and of the Wilson loop observable that
we study. 
By the well-known arguments, see e.g. \cite{Witten:1988ze,Cordes:1994sd} for a
general review and \cite{Pestun:2007rz} for the technical details on applying 
localization to the 1/2-BPS supersymmetric circular Wilson loops, the theory localizes
to the supersymmetric configurations $Q \Psi = 0$, where $\Psi$ denotes
fermionic fields of the theory. One can explain localization by deforming the
action of the theory by $Q$-exact term $S_{YM} \to S(t) = S_{YM} + t Q V$ with
$V = ( \Psi, \overline{Q \Psi})$ and sending $t$ to infinity. Since the bosonic part of the deformed
action is $S_{YM}^{bos} + t |Q \Psi|^2$, at the $t = +\infty$ limit the term $t
|Q \Psi|^2$ dominates. So, at the $t = +\infty$ limit, in the path integral we shall integrate only over
configurations solving $Q \Psi =0$ with the measure coming from the one-loop
determinant. On the other hand, the partition function and the expectation value
of observables do not depend on the $t$-deformation. Indeed, let the partition
function be  $Z(t) = \int e^{S(t)}$. Then, if $S(t)$ is
$Q$-closed and $\partial_{t} S(t)$ is $Q$-exact, we can integrate by parts in
$\p_t Z(t)$.  If the space of fields is essentially compact (all fields decrease
sufficiently fast at infinity) the boundary term vanishes and we obtain $\p_{t}
Z(t) = 0$.

In the present situation we use $V = (\Psi, \overline {Q \Psi})$. We recall, that
$\Psi$ is fermion of $\CalN=4$ super Yang-Mills obtained by dimensional
reduction of chiral sixteen-component spinor transforming in the
$S^{+}$ irreducible spin representation of  $\Spin(10)$. The other irreducible spin representation
$S^{-}$ of $\Spin(10)$ is dual to $S^{+}$. Therefore, there is a natural pairing
$S^{+} \otimes S^{-} \to \BC$, so that if $\psi \in S^{+}$ and $\chi \in S^{-}$
are spinors of the opposite chirality, the bilinear $(\chi, \psi)$ is $\Spin(10)$-invariant. (In components $(\chi,\psi)$ should be
read as $\sum_{\alpha=1}^{16} \chi_{\alpha} \psi_{\alpha}$ with no complex
conjugation operations).

In the Euclidean signature the representations $S^{+}$
and $S^{-}$ of $\Spin(10,\BR)$ are unitary and  complex conjugate to each other.
Hence, if $\chi \in S^{+}$ and $\psi \in S^{+}$ are spinors of the same
chirality, the bilinear $(\bar \chi,
\psi) = \sum_{\alpha=1}^{16} \overline{\chi}_{\alpha} \psi_{\alpha} $ is
invariant under $\Spin(10,\BR)$. So, because of our choice of Hermitian $Q$
(\ref{eq:generator-Q}) and because $Q$ squares to unitary global transformation
in $\SO(2)_S \times SO(2)_B$, the deformation term $V = (\overline \Psi,
Q\Psi)$ is $Q^2$-invariant and can be used for the localization.

The localization from the four-dimensional $\CalN=4$ SYM on $S^4$ to
a two-dimensional theory  on $\Sigma \subset S^4$ is done essentially in two steps.
It is convenient to represent the $S^4$ as an $S^2 \times S^1$ warped fibration over an interval
$I$ as in (\ref{eq:metric-polar-coordinates-S4}).

\begin{enumerate}
\item
We argue that $Q\Psi=0$ implies the invariance 
under the $SO(2)_S$ rotation, which acts by translation along the
$S^1$ fiber:  $\tau \to \tau + \const$. Hence, the $\CalN=4$ SYM  on $S^4$, for our purposes,
reduces to a three-dimensional theory on the manifold $D^3$ represented as a warped
$S^2$ fibration over $I$.

The resulting three-dimensional theory on $D^3$ 
can be interpreted as a deformed  version of certain cohomological field theory for extended Bogomolny
equations which were introduced by Kapustin and Witten in \cite{Kapustin:2006pk}. The interesting observables,
i.e. the Wilson loops (\ref{eq:Wilson-loops-S3}), are located at the boundary
$\Sigma = \p D^3$.

\item 

We show that physical action $S_{YM}$ for the reduced three-dimensional theory on $D^3$ can be
represented as a total derivative term modulo the equations $Q \Psi =
0$. Therefore, at the supersymmetric configurations $Q \Psi = 0$, 
the value of the reduced physical action $S_{YM}$ is determined by the boundary conditions at the $\Sigma$. 
The integral over the configurations satisfying $Q\Psi = 0$ reduces to an integral over
the boundary conditions on $\Sigma$.

\end{enumerate}

This is essentially the way how the two-dimensional theory appears. It turns out that
the resulting two-dimensional theory is closely related to topological
Higgs-Yang-Mills (or Hitchin-Yang-Mills) theory on $\Sigma$ studied in \cite{Moore:1997dj,Gerasimov:2006zt,Gerasimov:2007ap}.

 It is possible to introduce point-like singularities to solutions of the reduced equations
for the three-dimension theory on $D^3$ similar to the constructions in work~\cite{Kapustin:2006pk} by
 Kapustin and Witten. 
Such point singularities in the reduced three-dimensional 
theory on $D^3$ are uplifted to the codimension one singularities in the four-dimensional 
theory on $S^4$ and they are precisely the conformal supersymmetric 't Hooft operators 
as explained in \cite{Kapustin:2005py,Kapustin:2006pk}.

In this paper we do not consider the equations with singularities and 't Hooft 
operators. We study correlation functions only for  the initially introduced Wilson operators. 
However, we remark that  our construction in principle might be used to study correlation functions 
of a set of Wilson operators on $S^2$ and a set of 't Hooft operators, which are 
located on the $\U(1)$ orbits linking with the $S^2$. 

Also, it is possible to introduce codimension two singularity on the boundary of $D^3$. 
Such singularity corresponds to the disorder surface operator~\cite{Gukov:2008sn}
inserted on the two-sphere $S^2 = \p B^3$. This situation would 
be similar to the one studied in~\cite{Drukker:2008wr}. 
Again, in this work we aim to compute the expectation value only of Wilson loop operators on $S^2$
in absence of any extra singularities. We require all fields to be smooth and finite
in the path integral.

Now we give more details on the geometry of our setup.  The metric on $D^3$ in the first step above is 
\begin{equation}
  \label{eq:metric-on-B3}
  ds^2 = r^2 (d \xi^2 + \sin \xi^2 d\Omega_2^2) \quad \text{where} \quad 0 \leq  \xi \leq \pi/2,
\end{equation}
which is the round metric on a half of a three-sphere. 
Topologically $D^3 = (S^4 \setminus \Sigma)/SO(2)_S$ is a solid three-dimensional
ball. Under the $S^1$-fiber-forgetful projection $\pi: S^4 \to D^3$  the $\Sigma \subset S^4$,
where the interesting Wilson loops live, is mapped to the boundary of $D^3$. 
This boundary is located at $\xi = \pi/2$. 
The $S^1$ fiber shrinks to zero  at $\Sigma$.

\subsection{The supersymmetry equations}

\subsubsection{Choice of coordinates}
\newcommand{\x}{\tilde x}

To make the $\SO(2)_S \times \SO(3)_S$ isometry group of the $S^4$ explicit, we 
represent the metric as a warped product of the three-dimensional ball $D^3$ and the circle $S^1$. 
 On the $D^3$ we introduce the $\BR^3$
stereographic projection coordinates $\x_i$, and we keep the notation $\tau$ for
the coordinate on $S^1$. The metric in coordinates
$\x_i,\tau$  is then
\begin{equation}
  \label{eq:S^4-metric-B3-x-coordinates}
ds^2 (S^4) =  ds^2 (D^3 \times_w S^1) =  \frac {d\x_i d\x_i} { (1 + \frac {\x^2} {4r^2})^2} +  r^2 \frac { (1 - \frac{\x^2}{4 r^2})^2 }{(1  + \frac{\x^2}{4r^2})^2} d
\tau^2 \quad i=2,3,4
\end{equation}

One can write $S^4 = D^3 \times_w S^1$ where $w(\x)$ is the warp function $w(\x) = r^2 \cos^2 \xi = r^2
(1-\x^2/(4r^2))^2/(1+\x^2/(4r^2))^2$. The metric on $D^3$ is the standard round metric on the three-dimensional 
sphere.

The $\BR^4$ stereographic coordinates $x_i$  ($i=1\dots 4$)
and the $D^3 \times_w S^1$ coordinates $(\tau, \x_i)$  ($i=2,3,4$) are related
in a simple way. At the slice $x_1
= \tau = 0$ we have $x_i = \x_i$  ($i=2,3,4$). The generic relation between $x_i$ and
$(\tau, \x_i)$ is the following. From (\ref{eq:X-x-coordinates-relations})
on gets 
\begin{equation}
\label{eq:x-i-in-X_i}
  x_i = \frac {2} {1 + X_5/r} X_i, \quad i = 1\dots 4
\end{equation}
The $\SO(2)_S$ orbits are labelled by $(X_2,X_3,X_4)$. The $\tau$ is the
coordinate along $\SO(2)_S$ orbits, and we have \myref{2008-05-07 p3}
\begin{equation}
  \begin{aligned}
  X_1 &= R \sin \theta \sin \tau \\
  X_5 &= R \sin \theta \cos \tau.
  \end{aligned}
\end{equation}
So, from (\ref{eq:x-i-in-X_i}) we get the $\SO(2)_s$ orbits in the $\BR^4$
coordinates $x_i$, and hence, the transformation from coordinates $(\tau, \x_i)$ to
coordinates $(x_1, x_i)$
\begin{equation}
\begin{aligned}
\label{eq:x-xtilde-relation}
  x_i(\tau, \tilde x_i ) &= \tilde x_i \frac {1 + \sin \theta} { 1+ \sin \theta \cos \tau}, \quad \quad i = 2, \dots, 4 \\
  x_1(\tau, \tilde x_i) &=  R \frac { 2 \sin \theta \sin \tau} {1 + \sin \theta \cos \tau}
\end{aligned}
\end{equation}
where
\begin{equation}
  \label{eq:sin-theta-in-x}
  \sin \theta = \frac {1 - \frac{\x^2}{4 r^2}} { 1 + \frac{\x^2}{4 r^2}}.
\end{equation}
These $\SO(2)_S$ orbits are the usual circles in the $\BR^4$ coordinates $x_i$. These 
circles link with the two-sphere $S^2 = \{x_i | x_2^2 + x_3^2 +x_4^2 = 4 r^2,
x_1=0\}$ and are labeled by  points on $ D^3 = \{\x_i, \x^2 < 4r^2\}$.
For each $\x_i$ the corresponding circle is located in the two-plane
spanned by the vector $(1,0,0,0)$ and the vector $(0,\x_2,\x_3,\x_4)$.
The distance from the origin to the nearest point of the circle is $|\x|$, the
distance to the furthest point is $\frac {4r^2}{\x^2}$, while its center has coordinates
 $x_1 = 0, x_i = \tilde x_i (\frac 1 2 + \frac {r^2}{\x^2})$, and its diameter is $(4r^2 - \x^2)/|\x|$.

\subsubsection{Weyl invariance}

The supersymmetry equations $Q\Psi = 0$ are Weyl invariant. Indeed, given that under Weyl
transformation of metric $g_{\mu \nu}\to e^{2\Omega} g_{\mu \nu}$ the bosonic fields
transform as $A_{\mu} \to A_{\mu}, \Phi_{A} \to \Phi_{A} e^{-\Omega}, K_i \to K_i e^{-2\Omega}$ and the conformal Killing spinor transform as $\ve \to
e^{\frac 1 2 \Omega}\ve$, one gets that $Q_{\ve} \Psi \to e^{-\frac 3 2 \Omega} Q_{\ve}
\Psi$ which is a correct dimension for fermions. Therefore, the
localization procedure is essentially the same for two theories defined with
respect to the metrics related by a smooth Weyl transformation. (We ask transformation to
be smooth so that no conformal anomaly related to the infinity can appear.)

\newcommand{\w}{\tilde w}

In the coordinates $(\tau, \tilde x_i)$ the  $\SO(2)_{S} \times
\SO(3)_{S}$ symmetry is simply represented, so we shall start from the metric in the form
(\ref{eq:S^4-metric-B3-x-coordinates}).
Since $\x$ is bounded $|\x| < 2r$, the scale factor $(1+\x^2/(4r^2))$ is non-zero
and smooth everywhere over the $D^3$. It is convenient to get rid of this factor
in the equations by making Weyl transformation of the metric
$g_{\mu \nu} \to \tilde g_{\mu \nu} = (1+ \x^2/(4r^2))^2 g_{\mu \nu}$.
Under such rescaling the round spherical metric on $D^3$ becomes a flat metric. 
We refer to the $D^3$ equipped with the flat metric as the flat ball $B^3$.

So we study the equations $Q\Psi = 0$ on the space $B^3 \times_{\w} S^1$. The metric 
on this space is 
\begin{equation}
  \label{eq:metric-R3-tau}
  ds^2(B^3 \times_{\w} S^1) = d\x_i d\x_i + r^2 \left (1- \frac { \x^2}{4
      r^2} \right)^2 d \tau^2 \quad \text{where} \quad \x_i^2 \leq 4 r^2.
\end{equation}
We still call the coordinates on $B^3 \subset \BR^3$ as $\x_i$,
and the coordinate on $S^1$ as $\tau$, with the new warp factor being 
\begin{equation}
  \label{eq:warp-factor}
  \w(x) = r \lb 1- \frac {\x^2} { 4r^2} \rb.
\end{equation}
For fermions we use the following vielbein as an orthonormal basis in the cotangent bundle
\begin{equation}
  \label{eq:vielbein}
  (e_{i}) = (\w(x) d\tau, d\x_i), \quad i = 1\dots 4.
\end{equation}

\subsubsection{The diagonal $U(1) \subset \SO(2)_S \times \SO(2)_B$ invariance}

At $\tau = 0$ the coordinates $\x_{i}$ and corresponding vielbein coincide with
coordinates $x_i$. 
We take the conformal Killing spinor $\ve$ on the $B^3$
\begin{equation}
  \label{eq:conformal-Killing-spinor-inR3-metric}
  \ve(\x, \tau =0) = \hat \ve_{s} + \x^{i} \Gamma_{i} \hat \ve_{c}
\end{equation}
to write the supersymmetry equations at $\tau = 0$. Then, of course, using the $U(1) \subset \SO(2)_S \times \SO(2)_B$
invariance one can continue the equations to an arbitrary $\tau$. The Killing spinor $\ve$ on the whole
space $B^3 \times_{\w} S^1$ is invariant under the diagonal
$U(1) \subset \SO(2)_S \times \SO(2)_B$, i.e. under
simultaneous rotation of the $(X_5, X_1)$ and the $(\Phi_5, \Phi_0)$ planes.
A convenient change of variables for this diagonal $U(1)$ symmetry is to define 
the pair of ``twisted'' scalar fields\footnote{We remark that we are not making topological twisting
of the theory. All computations are done for the usual physical $\CalN=4$ SYM. We change variables
for a convenience but we do not change the Lagrangian and the observables.}     \myref{2008-05-08 p6}

\newcommand{\T}{T}
\newcommand{\R}{R}

\begin{equation}
\label{eq:change-of-variables-for-scalars-Phi5-Phi0}
\begin{aligned}
  \Phi_\T &= \cos \tau \Phi_0 - \sin \tau \Phi_5 \\
  \Phi_\R &= \sin \tau \Phi_0 + \cos \tau \Phi_5.
\end{aligned}
\end{equation}

\subsubsection{Conformal Killing spinor}
The conformal Killing spinor $\ve$ satisfies equation
\begin{equation}
  \label{eq:epsilon-tilde-is-derivative-of-epsilon}
  \nabla_{\mu} \ve = \Gamma_{\mu} \tilde \ve,
\end{equation}
and the off-shell transformation of fermions is given by
\begin{equation}
  \label{eq:susy-fermions}
  Q \Psi = \frac 1 2 F_{MN} \Gamma^{MN} \ve - 2 \Phi_A \tilde \Gamma^{A} \tilde
  \ve + i \nu_i K_i.
\end{equation}

The $\ve$ in components has explicit form
\begin{equation}
  \label{eq:epsilon-in-components}
  \ve =
  \begin{pmatrix}
    1  \\ 0 \\ 0 \\ 0
  \end{pmatrix} \otimes
  \begin{pmatrix}
    1  \\ 0 \\ 0  \\ 0
  \end{pmatrix}
- \frac 1 {2r}
\begin{pmatrix}
  0  \\ i \\ 0 \\0
\end{pmatrix} \otimes
\begin{pmatrix}
  0 \\ \x_2 \\ \x_3 \\ \x_4
\end{pmatrix}
\end{equation}
and $\tilde \ve$ is
\begin{equation}
  \label{eq:epsilon-tilde}
  \tilde \ve = \frac {1} {2r}
  \begin{pmatrix}
     0 \\ 0 \\ 0 \\ i
  \end{pmatrix} \otimes
  \begin{pmatrix}
     1 \\ 0 \\ 0 \\ 0
  \end{pmatrix}.
\end{equation}

\subsubsection{Off-shell closure}

We also  need 7 auxiliary spinors $\nu_i$ to write down the off-shell
closure of the supersymmetry transformations (\ref{eq:susy-fermions}) like in
\cite{Baulieu:2007ew,Berkovits:1993zz}. It is easy to find such set of $\nu_i$
because only top 8 components of $\ve$ are non-zero. More invariantly, $\ve$
satisfies
\begin{equation}
  \label{eq:constraint-on-ve-chirality}
  (\Gamma^{1} + i \Gamma^{0}) \ve = 0,
\end{equation}
i.e. it is chiral with respect to the $SO(8)$ acting on the vector indices
$2,\dots,9$. Then, as a set of 7 spinors $\nu_i$, one can choose
\begin{equation}
  \label{eq:nu-i}
  \nu_i = \Gamma_{9 i} \ve \quad \text{for} \quad i = 2, \dots 8.
\end{equation}
Such spinors $\nu_i$ are again $\SO(8)$ chiral and 
have only $8$ top components being non-zero.

\subsubsection{Splitting of the supersymmetry equations: top and bottom}

To compute the components of  $Q\Psi$  it is convenient to split sixteen component
spinors in $S^{+}$ into two eight-component spinors on which $\Gamma^{1}\Gamma^{0}$ acts by $+i$ or
$-i$ respectively. (We will use interchangeably space-time index $1$ or $\tau$ to
denote direction along the coordinate $\tau$ in (\ref{eq:metric-R3-tau}).)
With our choice of gamma-matrices (\ref{eq: gamma-matrices}), 
if the eight-component spinors are called $\Psi^{t}$ and $\Psi^{b}$, we have 
\begin{equation}
  \label{eq:psi-in-terms-psi-t-psi-b}
  \Psi =
\begin{pmatrix}
    \Psi^{t} \\
    \Psi^{b}
  \end{pmatrix},
\end{equation}
and  
\begin{equation}
  \begin{aligned}
  \label{eq:epsilon-splitting}
  \ve &=
  \begin{pmatrix}
    \ve^{t} \\
    0
  \end{pmatrix} &
  \tilde \ve &=
  \begin{pmatrix}
    0 \\
    \tilde \ve^{b}
  \end{pmatrix}.
  \end{aligned}
\end{equation}
Next, we represent the eight-component spinors
$\Psi^{t}$ and $\Psi^{b}$ by the octonions $\BO$. A spinor
\begin{equation}
  \label{eq:psi-in-octonions}
  \Psi^t =
  \begin{pmatrix}
    \Psi_{1} \\
    \Psi_{2} \\
    \dots \\
    \Psi_{8}
  \end{pmatrix}
\end{equation}
will be written as 
\begin{equation}
  \label{eq:psi-in-octonions-presented}
  \Psi^{t} = \Psi^t_{1} e_9 + \Psi^t_{2} e_2 + \dots + \Psi^t_{8} e_8,
\end{equation}
where $e_9,e_2, \dots, e_8$ are the basis elements of $\BO$, see explanation
after (\ref{eq: gamma-matrices}).
Similarly,
\newcommand{\e}{\tilde e}
\begin{equation}
  \label{eq:psi-in-octonions-presented-bottom}
  \Psi^{b} = \Psi^b_{1} \e_9 + \Psi^b_{2} \e_2 + \dots + \Psi^b_{8} \e_8,
\end{equation}
where $\e_9,\e_2,\dots,\e_8$ are the basis elements in the second copy of $\BO$
representing the bottom components of $\Psi$. In these notations
\begin{equation}
  \label{eq:epsilon-in-octonions}
  \ve = e_9 - \frac {i} {2r} \x_i e_{i+4}
\end{equation}
and
\begin{equation}
  \label{eq:epsilon-tilde-in-octonions}
  \tilde \ve = \frac {i} {2r} \e_{5}.
\end{equation}

\subsection{Bottom equations and the circle invariance}
Now we analyze the bottom components of the equations
(\ref{eq:susy-fermions}).

Taking into account the chiral structure of gamma-matrices (\ref{eq:
  gamma-matrices}) and spinors $\ve, \tilde \ve$ as in
(\ref{eq:epsilon-splitting}), we get
\begin{multline}
\label{eq:QPsib}
  Q\Psi^{b} = \sum_{m=\hat 2\dots \hat 9} (F_{\hat 0 \hat m} \Gamma^{\hat 0 \hat
    m} + F_{\hat 1 \hat m} \Gamma^{\hat 1 \hat m}) \ve - 2 \Phi_0 \tilde
  \Gamma^{0} \tilde \ve  = \\
 - ( i F_{\hat{0} \hat{m}} + F_{\hat {1} \hat{m}}) E_{\hat m} \ve + 2 i \Phi_0
 \tilde \ve =   - ( i F_{\hat{0} \hat{m}} + F_{\hat {1} \hat{m}}) e_{\hat m} (
 e_9 - \frac {i} {2r} \x_i e_{i+4} ) + 2 i \Phi_0 \frac {i} {2r} \e_{5}.
\end{multline}
We use indices with  hat to denote vector components with respect to the
orthonormal vielbein (\ref{eq:vielbein}), e.g. $F_{\hat 1  \hat m} = \w(x)^{-1} F_{\tau \hat m}$.
For brevity we consider equations along the radial line $(\tau, \x) =
(0,\x_2, 0, 0)$, and then, using the $\SO(2)_S$ and the $\SO(3)_S$ symmetry we can write the equations 
on the the whole space $B^3 \times_{\w} S^1$.
At $\x_2 < 2r$, the six equations corresponding to the components $\hat m = 3,4,6,7,8,9$
are linearly independent \myref{2008-05-08 p4} and imply
\begin{equation}
  \label{eq:vanishing-F_01-six-equations}
  i F_{\hat 0 \hat m} + F_{\hat 1 \hat m} = 0 \quad \text{for} \quad \hat m = 3,4,6,7,8,9.
\end{equation}

We can make diagonal transformation in $\SO(2)_S \times \SO(2)_B$ like in
(\ref{eq:change-of-variables-for-scalars-Phi5-Phi0})
to transform (\ref{eq:vanishing-F_01-six-equations}) to an arbitrary $\tau$
\begin{equation}
  \label{eq:arbitrary-tau}
  i F_{\hat m \T} + \frac {1} {r(1 - \frac {\x^2} {4r^2})} F_{\hat m \tau} = 0 \quad \hat m
  = 3,4,6,7,8,9
\end{equation}
where we replaced index $\hat 1$ by $\tau$ using the scaling function $\w(\x)$,
and where $F_{\T \hat m} =[\Phi_\T, \nabla_{\hat m}] = -\nabla_{\hat m} \Phi_T$.
Next we consider the remaining two components in (\ref{eq:QPsib}) for the basis
elements $e_2$ and $e_5$. At $\tau = 0$ we have
\begin{equation}
\begin{aligned}
  iF_{\hat 0 \hat 2} + F_{\hat 1 \hat 2} - \frac {i} {2r} \x_2 (i F_{\hat 0
    \hat 5} + F_{\hat 1 \hat 5})
  & = 0   \quad \text{(on $e_2$)} \\
  iF_{\hat 0 \hat 5} + F_{\hat 1 \hat 5} + \frac {i} {2r} \x_2 (i F_{\hat 0 \hat
    2} +
  F_{\hat 1 \hat 2}) - \frac 1 r \Phi_0 & = 0  \quad \text{(on $e_5$)}.
\end{aligned}
\end{equation}
Again we can make $\tau$ arbitrary by making the diagonal transformation $\U(1) \in \SO(2)_S
\times \SO(2)_B$
\begin{equation}
\label{eq:FTR-equations}
  \begin{aligned}
  (i F_{\T \hat 2} + \w^{-1} F_{\tau \hat 2}) - \frac {i} {2r} \x_2 (i F_{\T \R} +
  \w^{-1} (F_{\tau \R} - \Phi_\T) )  &= 0 \\
  (i F_{\T \R} + \w^{-1} (F_{\tau \R} - \Phi_\T)) + \frac {i}{2r} \x_2 ( i F_{\T 2} +
  \w^{-1} F_{\tau 2}) + \frac {1} {r} \Phi_\T &= 0.
  \end{aligned}
\end{equation}
The first line plus the second multiplied by $i \x_2 /2r$ is
\begin{equation}
\label{eq:phiT-equation}
  i ( 1-\frac {\x^2}{4r^2}) F_{\T 2} + \frac 1 r F_{\tau 2} + i \frac {\x_2}{2
    r^2} \Phi_\T = 0.
\end{equation}
Introducing a rescaled field
\begin{equation}
  \label{eq:rescaled-Phi-T}
  \tilde \Phi_\T = r (1 - \frac {\x^2}{4r^2}) \Phi_\T,
\end{equation}
the equation (\ref{eq:rescaled-Phi-T}) is rewritten as
\begin{equation}
  i \nabla_{2} \tilde \Phi_{\T} +  F_{2 \tau}  = 0.
\end{equation}
The remaining equation from (\ref{eq:FTR-equations}) is then
\begin{equation}
\label{eq:phiTR}
  i (1 - \frac {\x^2}{4r^2}) F_{\T \R} + \frac {1} r F_{\tau \R} = 0.
\end{equation}
We can summarize the 8 equations
(\ref{eq:arbitrary-tau}),(\ref{eq:rescaled-Phi-T}),(\ref{eq:phiTR}) resulting from $Q\Psi^{b} = 0$:
\begin{equation}
\label{eq:summmary-QPsi-bottom}
  [\nabla_{\hat m}, \nabla_{\tau} + i  \tilde \Phi_\T ]  = 0 \quad
 \text{for} \quad \hat m = 2,3,4,R,6,7,8,9.
\end{equation}
One can introduce complexified connection $\nabla_{\tau}^{\BC} = \nabla_{\tau} +
i \tilde \Phi_\T$ and interpret the equations (\ref{eq:summmary-QPsi-bottom}),
as vanishing of the electric field (the three equations $F^{\BC}_{\tau \hat i} = 0$, $i =
2,3,4$) and covariant time independence of the remaining five scalars
($\nabla_{\tau}^{\BC} \Phi_{\R,6,7,8,9}=0$) in the conventions where $\tau$ is the time
coordinate.

Since $Q^2$ generates  translations along $\tau$, we can interpret $Q^2$ as the
Hamiltonian. 
The bottom equations (\ref{eq:summmary-QPsi-bottom}) say that momenta of
all fields vanish and that the theory localizes to some three-dimensional
theory.
This three-dimensional theory is defined on a three-dimensional ball $B^3$ whose
boundary is the two-sphere $\Sigma$ where interesting Wilson loop operators are
located.

The supersymmetric configurations in this three-dimensional theory are
determined by the top eight components of the equations $Q\Psi = 0$, which we 
shall analyze now. 

\subsection{Top equations and the three-dimensional theory }
Writing the top eight components of  $Q\Psi$ explicitly we get \myref{2008-05-12 p1}
\begin{multline}
  \label{eq:top-eight}
  Q \Psi^{t} = F_{\hat 0 \hat 1}\Gamma^{\hat 0 \hat 1}\ve^{t} + \sum_{2 \leq m < n \leq
    9} F_{m n} \Gamma^{mn} \ve^{t} - 2 \tilde E_A \Phi_A \tilde \ve^{b} +
  \sum_{1 \leq I \leq 8} i K_I
    \Gamma^{9I} \ve^t  = \\
= -i F_{\hat 0 \hat 1} \ve^{t} + (F_{9I} + i K_I)E_{I} \ve^t - \sum_{2 \leq I < J
  \leq 8} F_{IJ}
E_I E_J \ve^t - 2 \tilde E_{A}\Phi_A \tilde \ve^b.
\end{multline}
In the following we use indices $I,J=2,\dots, 8$ and $i,j,k,p,q = 2, \dots,
4$. In this section we put $r=1/2$ to avoid extra factors. We  do not write
tilde over $x$ understanding that $x^{i}$  ($i=2,3,4$) are the coordinates on
the flat unit ball $B^3
\subset \BR^3$. The antisymmetric symbol $\ep_{ijk}$ is defined as $\ep_{234} = 1$.
The following multiplication table of octonions is helpful
\begin{equation}
\begin{aligned}
  e_i e_j &= \ep_{ijk} e_k - \delta_{ij} e_9 \\
  e_{i+4} e_i &= e_5 &
      e_i e_5 &= e_{i+4} &
   e_5 e_{i+4}&= e_i \\
  e_k e_{i+4} &= -\ep_{kij}e_{j+4} - \delta_{ik} e_5 &
  e_{i+4} e_{j+4} &= - \ep_{ijk} e_k - \delta_{ij} e_9 &
  e_{j+4} e_k &= \delta_{jk}e_5 - \ep_{jki} e_{i+4}
\end{aligned}
\end{equation}

After some algebra \myref{2008-05-14 p3} we get
the first term
\begin{equation}
  \label{eq:top-part1}
Q\Psi^{t(1)} =  -i F_{\hat 0 \hat 1} \ve = -iF_{\hat 0 \hat 1}(e_9 - i x_j e_{j+4}),
\end{equation}
the second term
\begin{multline}
Q\Psi^{t(2)}= (F_{9I} + i K_I)E_{I} \ve = (F_{9I}+ i K_I)e_I(e_9 - i x_j e_{j+4}) = \\
(F_{9i} + i K_i)(e_i + i x^j \ep_{ijk} e_{k+4} + i x^j \delta_{ij} e_5) + \\
(F_{95} + i K_5)(e_5 - i x_j e_j) + \\
(F_{9\,i+4} + i K_{i+4})(e_{i+4} + i x^j \ep_{ijk} e_k + i x^j \delta_{ij} e_9),
\end{multline}
the third term
\begin{multline}
Q \Psi^{t(3)} = -F_{I < J} E_I E_J \ve = \\
=\left[-\frac 1 2 (F_{ij} - F_{i+4 \, j+4})\ep_{ijk} e_k + F_{i \,
  j+4} \ep_{ijk} e_{k+4} + F_{i\, i+4} e_5 - F_{5 \, k+4} e_k
- F_{k5} e_{k+4} \right] \\
+ i \left[
% x^k F_{i < j}(-\ep_{jkp})(-\ep_{ipq} e_{q+4} - \delta_{ip} e_5) + x^k
%   F_{i < j}(-\delta_{jk}) e_{i+4}  \\
% + x^k F_{i5}(\ep_{ikp} e_p - \delta_{ik} e_9) \\
% + x^k F_{i \,j +4} (-\ep_{jkp})(\ep_{ipq} e_{q} - \delta_{ip} e_9) + x^k F_{i \,
%   j+4} e_i (-\delta_{jk})  \\
% + x^k F_{5 j+4}(-\ep_{jkp})(-e_{p+4}) + x^k F_{5 j+4}(-\delta_{jk})e_5 \\
% + x^k F_{i + 4 < j + 4}(-\ep_{jkp})(-\ep_{ipq} e_{q+4} + \delta_{ip} e_5) + x^k
% F_{i+4 j+4} (-\delta_{jk}) e_{i+4} \left.
% in \myref{2008-05-14 p3} it is simplified to
F_{ij} x_i e_{j+4} + \frac 1 2 F_{ij} x_k \ep_{ijk} e_5 \right. \\
+ F_{i5} x_k \ep_{ikj} e_j - F_{i5} x_i e_9
- F_{i \, j+4} x_i e_j - F_{i \, j+4} x_j e_i + F_{i \, i+4} x_k e_k + F_{i \,
  j+4} x_k \ep_{ijk} e_9 \\
+ F_{5, i+4} x_k \ep_{ikj} e_{j+4} - F_{5 j+4} x_j e_5 +
 \left. F_{i+4 \, j+4} x_i e_{j+4} - \frac 1 2 F_{i+4 j+4} \ep_{ijk} x_k e_5
  \right ]
\end{multline}
and the fourth term
\begin{equation}
  \label{eq:Q-top-4-terms}
  Q\Psi^{t(4)} = - 2 \tilde E_{A}  \Phi_A \tilde \ve^{b} = -2i ( \Phi_9 e_5 +
  \Phi_5 e_9 + \Phi_{i+4} e_i).
\end{equation}

Now we analyze the equations. We have eight complex, i.e. sixteen real, equations on eight real physical
fields $A_{2,3,4},\Phi_{\R,6,7,8,9}$ and seven real auxiliary fields $K_i$. 
(The deformation term  $t |Q\Psi|^2$ vanishes on
the real integration contour if and only if both imaginary and complex part of $Q \Psi$ 
vanishes.) 
%Hence, the top equations $Q \Psi^{t}=0$ naively imply 16 real equations.
We shall see shortly that only 15 equations are independent. Seven
auxiliary fields can be easily integrated out. Then we are left with eight equations. One of these eight equations
gives real constraint on the complexified time connection: \myref{2008-05-14}
\begin{equation}
  \label{eq:QPsi-top-tau-T}
  [\nabla_{\tau},\tilde \Phi_{\T}] = 0.
\end{equation}
This equation together with (\ref{eq:summmary-QPsi-bottom}) completes our claim that the field configurations are all $\tau$-invariant up to a gauge transformation.

What remains is the system of seven first order differential equations in
three dimensional space on gauge field and
five scalars. The equations are gauge invariant. Modulo gauge transformations, the system is elliptic in the
interior of the three-dimensional ball $B^3$. The system is closely related to the extended
three-dimensional Bogomolny equations studied in \cite{Kapustin:2005py}.

Now we shall give technical details on the equations.
First we eliminate $\Im Q\Psi^t |e_{9}$ by adding to it $-x_i \Re  Q\Psi
|_{e_{i+4}}$ \myref{2008-05-14 p4}
\begin{multline}
\label{eq:PHI5-constant}
  \Im Q\Psi |e_{9} - x_i \Re Q \Psi |_{e_{i+4}} =
-F_{\hat 0 \hat 1} + F_{9 \, i+4} x_i -F_{i5} x_i + F_{i j+4} x_k \ep_{ijk} - 2
\Phi_5  \\ - (-F_{\hat 0 \hat 1} x^2 + F_{9 \, i+4} x_i - F_{i5} x_i + F_{i \, j+4}
x_k \ep_{ijk}) = \\
= -F_{\hat 0 \hat 1} (1 - x^2) - 2 \Phi_5  = 2 [\nabla_{\tau} \Phi_{T}]
\end{multline}
This is the real equation which completes the system of time-invariance equations
(\ref{eq:summmary-QPsi-bottom}).

Next we consider $\Re Q \Psi^t |_{e_9}$:
\begin{equation}
\label{eq:extraK}
\Re Q \Psi^{t} |e_{9} = -K_{i+4} x_i
\end{equation}
This equation is one constraint on the auxiliary fields $K_i$.
We are left with 14 more equations $\Im Q\Psi^{t}|_{e_{I}} = 0$ and $\Re Q
\Psi^{t}|_{e_{I}} = 0$, $I=2,\dots,8$. Using $\Im Q\Psi^{t}|_{e_{I}} = 0$ we shall solve for
$K_I$ in terms of the physical fields $A$ and $\Phi$, and we will see actually that the constraint (\ref{eq:extraK}) is
automatically implied.

The seven equations $\Im Q\Psi^{t}|_{e_{I}} = 0$ imply \myref{2008-05-14 p7}
\begin{equation}
  \label{eq:K-fields-in-terms-of-APHI}
\begin{aligned}
K_{k} & = F_{95} x_k - F_{9 \, i+4} \ep_{ijk} x_j - F_{i5} x_j \ep_{ijk} + F_{i \,
  k+4} x_i + F_{k \,i+4} x_i - F_{i \, i+4} x_k + 2 \Phi_{k+4} \\
K_{5} & = - F_{9i} x_i - \frac 1 2 F_{ij} x_k \ep_{ijk} + F_{5 \, j+4} x_j +
\frac 1 2 F_{i +4 \, j+4} x_k \ep_{ijk} + 2 \Phi_9 \\
K_{k+4} &= - F_{9i} \ep_{ijk} x_j - F_{ik} x_i - F_{5 \, i+4} x_j \ep_{ijk} -
F_{i+4 \, k+4} x_i.
\end{aligned}
\end{equation}
The seven components $\Re Q\Psi^{t}|_{e_{I}} = 0$ are \myref{2008-05-14 p8}
\begin{equation}
  \label{eq:APHI-with-K-top-equations}
\begin{aligned}
 \Re Q \Psi^{t}|_{e_k}  &= F_{9k} - \frac 1 2 (F_{ij} - F_{i+4\, j+4})\ep_{ijk} -
 F_{5 \, k+4} + K_5 x_k - K_{i+4} x_j \ep_{ijk}  \\
 \Re Q \Psi^{t} |_{e_5} & = F_{95} + F_{i \, i+4} - K_{i} x_i  \\
 \Re Q \Psi^{t} |_{e_{k+4}} & = F_{9 \, k+4} + F_{i \, j+4} \ep_{ijk} - F_{k5} +
 2 \Phi_5 (1-x^2)^{-1} x_k - K_i x_j \ep_{ijk}.
\end{aligned}
\end{equation}

After plugging in (\ref{eq:APHI-with-K-top-equations}) the expressions for $K_I$
(\ref{eq:K-fields-in-terms-of-APHI}) we get \myref{2008-05-14 p 16}
\begin{equation}
  \label{eq:final-top-equations-original-variables}
\begin{aligned}
& \begin{split} \Re Q \Psi^{t}|_{e_k}  = F_{9k}(1-x^2) - \frac 1 2 F_{ij} \ep_{ijk} (1+x^2) + \frac 1 2
 F_{i+4\, j+4} \ep_{ijp}(\delta_{pk} - x^2 \delta_{pk} + 2 x_p x_k) - \\
 F_{5 \, j+4}(\delta_{jk} + x^2 \delta_{jk} - 2x_j x_k) + 2 \Phi_9 x_k
\end{split}\\
& \begin{split}
 \Re Q \Psi^{t} |_{e_5}  = F_{95}(1-x^2) + F_{i \, j+4}(\delta_{ij} +
 \delta_{ij} x^2 - 2 x_i x_j) - 2 \Phi_{j+4} x_j
\end{split}\\
& \begin{split} \Re Q \Psi^{t} |_{e_{k+4}}  = F_{9 \, i+4}(\delta_{ik} + x_i x_k - x^2
 \delta_{ik})  - F_{i5}(\delta_{ik} - x_i x_k + x^2 \delta_{ik}) +
 2 \Phi_5 (1-x^2)^{-1} x_k \\ + F_{i \, j+4} ( \ep_{ijk} - x_i
 x_p \ep_{jpk} - x_j x_p \ep_{ipk}) - 2 \Phi_{i+4} \ep_{ijk} x_j e_{k+4}.
\end{split}
\end{aligned}
\end{equation}
The above calculations are done at the slice $\tau=0$. For an arbitrary $\tau$
the field $\Phi_5$ should be replaced by $\Phi_\R$ as in
(\ref{eq:change-of-variables-for-scalars-Phi5-Phi0}).

\subsubsection{Simplification at the origin: extended Bogomolny equations}

Let us analyze the equations $\Re Q \Psi^{t} |_{e_{I}} = 0$ using (\ref{eq:final-top-equations-original-variables}).
At $x_i=0$ the equations simplify to
\begin{gather}
\label{eq:extended-Bogomolny-1}
 -*(F - \Phi \wedge \Phi) - d_A \Phi_9 + [\Phi, \Phi_\R] = 0  \\
\label{eq:extended-Bogomolny-2}
* d_A \Phi - d_A \Phi_\R - [\Phi, \Phi_9] = 0 \\
\label{eq:extended-Bogomolny-3}
d_A *\Phi + [\Phi_9, \Phi_\R] = 0.
\end{gather}
where we identified the three scalar fields $\Phi_{i+4}$ with the components of
one-form $\Phi$ on $\BR^3$, we set $\Phi = \Phi_{i+4} dx^{i}$, and $*$ is the Hodge
operator on $\BR^3$ equipped with the standard flat metric.

Let us combine the gauge field $A$ and the one-form $\Phi$ into a complexified
connection $A_\BC = A + i \Phi$, and similarly combine the scalars $\Phi_\R$ and
$\Phi_9$ into complexified scalar $\Phi_{\BC} = \Phi_9 + i \Phi_\R$. 
%%%Important misprint is here
%%%$\Phi_\BC = \Phi_\R + i \Phi_9$. 
%%%

Then the
equations
(\ref{eq:extended-Bogomolny-1})(\ref{eq:extended-Bogomolny-2})
can be written as
\begin{gather}
  -*\Re F_\BC - \Re d_{A_\BC} \Phi_\BC = 0 \\
   *\Im F_\BC - \Im d_{A_\BC} \Phi_\BC = 0.
\end{gather}
This pair of real equations can be combined into one complex equation
\begin{equation}
\label{eq:extended-Bogomolny-complexifed}
   * \overline{F_{\BC}} + d_{A_\BC} \Phi_\BC  = 0.
\end{equation}
The equation (\ref{eq:extended-Bogomolny-complexifed}) was
introduced by Kapustin and Witten in \cite{Kapustin:2006pk} and is called \emph{extended Bogomolny equation}.

\subsubsection{The three-dimensional equations in rescaled variables}

Hence, we see that at the origin of $\BR^3$, the equations
(\ref{eq:final-top-equations-original-variables}) look exactly like the 
relatively familiar system of elliptic equations. Away from $x=0$ the equations are deformed into something
more complicated. We will try to make some simple rescaling of variables to
convert the equations to more standard form.

\newcommand{\Phit}{\tilde \Phi}
For this purpose, we rescale the scalar fields and define
$\Phit_{j}$, $j=2,3,4$, by 
\begin{equation}
  \label{eq:change-of-variables-PhiI4}
  \Phi_{i+4} = \Phit_{j}\left(\delta_{ij} + \frac{ 2 x_i x_j} {1 - x^2}\right).
\end{equation}
This change of variables is smooth in the interior of the ball $B^3$.
In terms of $\Phit_{i}$ the first equation in
(\ref{eq:final-top-equations-original-variables}) becomes \myref{2008-05-27 p1}
\begin{equation}
  \label{eq:my-first-equation-Phi-tilde}
  -\frac 1 2 (1+x^2) \ep_{ijk} (F_{ij} - [\Phit_{i}, \Phit_{j}]) -
  \nabla_{k}((1-x^2)\Phi_9) + (1+x^2)[\Phit_k, \Phi_\R] = 0.
\end{equation}
The second equation in (\ref{eq:final-top-equations-original-variables}) becomes
\myref{2008-05-27 p3}
\begin{equation}
  \label{eq:my-second-equation-Phi-tilde}
  (1-x^2) \ep_{ijk} \nabla_{i} \Phit_{j} - \nabla_k ((1-x^2)\Phi_\R)
 - \frac {1-x^2}{1+x^2}\left((1-x^2)\delta_{ik}+ \frac{ 4 x_i x_k}{1-x^2}\right) [\Phit_i,\Phit_9] =0.
\end{equation}
Finally, the third equation in (\ref{eq:final-top-equations-original-variables})
becomes \myref{2008-05-27 p5}
\begin{equation}
  \label{eq:my-third-equation-Phi-tilde}
  (1+x^2) \nabla_{i} \Phit_{i} + 2 \frac {3+x^2}{1-x^2} x_i \Phit_i + (1-x^2) [\Phi_9,
  \Phi_5] = 0.
\end{equation}

%the moduli space of solutions of SUSY equations with finite YM action
\newcommand{\M}{\CalM}

\subsubsection{The three-dimensional equations linearized}

Let $\CalM$ denote  the moduli space
of smooth solutions to 
(\ref{eq:my-first-equation-Phi-tilde}),(\ref{eq:my-second-equation-Phi-tilde}),(\ref{eq:my-third-equation-Phi-tilde})
with finite Yang-Mills action.
In the localization computation we need to integrate over $\CalM$. 
Clearly, the zero configuration $A=\tilde \Phi=0,
\Phi_\R = \Phi_9 = 0$ is a solution. Let us analyze the linearized problem near
the zero configuration, in other words, let us find the fiber of the tangent space
$T\CalM_0$.
The linearized equations
(\ref{eq:my-first-equation-Phi-tilde}),(\ref{eq:my-second-equation-Phi-tilde}),(\ref{eq:my-third-equation-Phi-tilde})
are
\begin{gather}
\label{eq:linearized-equations}
  (1+x^2) *_{\BR^3} dA + d((1-x^2)\Phi_9) = 0\\
\label{eq:linearized-equations2}
  (1-x^2) *_{\BR^3} d\Phit - d((1-x^2)\Phi_\R) = 0 \\
\label{eq:linearized-equations3}
  (1+x^2) d^{*}_{\BR^3} \Phit + 2 \frac {x^2 + 3}{1-x^2} (x,\Phit) =0.
\end{gather}
Here we by $*_{\BR^3}$ we denoted the Hodge star operation with respect to the
standard metric on $\BR^3$. It is possible to absorb extra $(1\pm x^2)$ factors in
the Hodge star operation using a rescaled metric.
We will use the metric
\begin{equation}
  \label{eq:metric-S3}
  ds^2(S^3) = \frac{dx_i dx_i}{(1 + x^2)^2}, \quad |x| < 1
\end{equation}
which is the metric on a half of the round  $S^3$,
and
\begin{equation}
  \label{eq:metric-H3}
  ds^2(H_3) = \frac{dx_i dx_i}{(1-x^2)^2}, \quad |x| < 1
\end{equation}
which is a metric on hyperbolic space $H^3$ in Poincare coordinates.
Then the first two equations in (\ref{eq:linearized-equations}) turn into
\begin{gather}
  \label{eq:linearizedA}
  *_{S^3} dA + d \tilde \Phi_9 = 0 \\
  \label{eq:linearizedPhi}
  *_{H^3} d \Phit  - d \tilde \Phi_\R = 0,
\end{gather}
where
\begin{align}
  \label{eq:Phi5-rescaled-definition}
  \tilde \Phi_\R = (1-x^2) \Phi_\R \\
  \label{eq:Phi9-rescaled-definition}
  \tilde \Phi_9 = (1-x^2) \Phi_9.
\end{align}
The  equation (\ref{eq:linearizedA}) implies that $\tilde \Phi_9$ is harmonic
for the $S^3$ metric
\begin{equation}
  \label{eq:laplacian-Phi9-zero}
 \Delta_{S^3} \tilde \Phi_9 = 0,
\end{equation}
and the equation (\ref{eq:linearizedPhi}) implies that $\tilde \Phi_5$ is
harmonic for the $H_3$ metric
\begin{equation}
  \label{eq:laplacian-Phi5-zero}
 \Delta_{H^3} \tilde \Phi_\R = 0.
\end{equation}
We need to consider only such solutions that the fields $\Phi_\R, \Phi_9$ are not singular at
the boundary. (Singular solutions can be considered too, but they correspond to
the disorder surface operator~\cite{Drukker:2008wr} inserted on the two-sphere $S^2 = \p B^3$. In this
work we aim to compute the expectation value of Wilson loop operators on $S^2$
in the absence of any surface operators. Hence we require $\Phi_\R$ and $\Phi_9$
fields to be finite at the $S^2$.)
If $\Phi_\R$ and $\Phi_9$ fields are finite at $|x|=1$, then
$\tilde \Phi_\R$ and $\tilde \Phi_9$ vanish there by
(\ref{eq:laplacian-Phi9-zero}),(\ref{eq:laplacian-Phi5-zero}).
Hence we have the Laplacian problem
(\ref{eq:laplacian-Phi9-zero})(\ref{eq:laplacian-Phi5-zero}) with Dirichlet
boundary conditions
\begin{equation}
  \label{eq:Dirichlet-boundary-conditions}
  \tilde \Phi_\R |_{\p B^3} = \tilde \Phi_9|_{\p B^3} = 0. \\
\end{equation}
Since a harmonic function $Y(x)$ vanishing on the boundary must vanish (it can be shown
integrating by parts $\int_B dY \wedge * dY =  \int_{\p B} Y \wedge * dY$), we conclude that there is no
nontrivial finite solution for the fields $\Phi_\R, \Phi_9$, so
\begin{equation}
  \label{eq:solution-Phi5-Phi9=0}
  \Phi_\R = \Phi_9 = 0.
\end{equation}

\paragraph{Explicit solutions in spherical harmonics.}

One might worry that this argument  fails for the $H^3$ because of 
the infinite boundary. However, the explicit solution of the Laplace
equation on  $H^3$  shows that all radial
wave-functions, which are smooth in the interior of $H^3$, do not vanish at the
boundary.
In spherical coordinates, the $H^3$ metric is
\begin{equation}
  \label{eq:H3-metric}
  ds^2 = \frac { d\xi^2 + \sin^2 \xi d\Omega_2^2}{\cos^2 \xi},
\end{equation}
where $\xi$ is the radial coordinate $0 \leq \xi < \pi/2$ and $d\Omega_2^2$ is
the standard metric on the unit two-sphere.
Then
\begin{equation}
  \label{eq:laplacian-on-H3}
  \Delta_{H^3} f  = \frac {1} {\sqrt{g}} \p_i (\sqrt{g} g^{ij} \p_j) f =
\frac {\cos^3 \xi}{\sin^2{\xi}} \p_{\xi} \left ( \frac {\sin^2 \xi}{\cos \xi} \p_{\xi}
f \right) + \frac {\cos^2 \xi}{\sin^2 \xi} \Delta_{S^2}f.
\end{equation}
If $f_s(\xi)$ is the radial wave-function for the angular momentum $s$ on the
$S^2$ then $\Delta_{S^2}f_s = -s(s+1) f_s$. So the equation
(\ref{eq:laplacian-on-H3})
is a special case of the Laplace equation in the $(p,q)$ polyspherical coordinates (see e.g. \cite{MR0229863} p.499)
\begin{equation}
  \label{eq:Laplace-equation-polyspherical-coordinates}
\begin{split}
 \frac {1} {\cos^{p} \xi  \sin^{q} \xi } \frac{\p}{\p \xi}\left(\cos^{p} \xi \sin^{q} \xi
  \frac {\p u} {\p \xi } \right) - \left (  \frac {r(r+p-1)}{\cos^2 \xi} + \frac
    {s(s+q -1)}{\sin^2 \xi} - l(l+p+q) \right) u = 0
\end{split}
\end{equation}
for $q=2, p=-1,r=0,l=0$.
The solutions of (\ref{eq:Laplace-equation-polyspherical-coordinates})
non-singular at $\xi=0$ are
\myref{2008-05-27 p4}
\begin{equation}
  u =  \tan^{s} \xi F \left( \frac {s - l + r}{2}, \frac { s - l - r - p + 1 } {2}, s +
  \frac {q + 1}{2}; -\tan^{2} \xi \right),
\end{equation}
where $F(\alpha,\beta,\gamma;z)$ is the $_2 F_1 $ hypergeometric function.
In our case we have
\begin{equation}
  \label{eq:solution-fs}
  f_s (\xi) = \tan^{s} \xi F (s/2, s/2 + 1, s+3/2, - \tan^{2} \xi).
\end{equation}
Using identity
\begin{equation}
  \label{eq:hyper-geometric-identity}
  F(\alpha,\beta,\gamma,z) = (1-z)^{-\alpha}F(\alpha,\gamma-\beta,\gamma;\frac{z}{z-1})
\end{equation}
we can rewrite (\ref{eq:solution-fs}) as
\begin{equation}
  \label{eq:solution-fs-sin-xi}
  f_s (\xi) = \sin^{s} \xi F(s/2, s/2 + 1/2, s+ 3/2, \sin^2 \xi).
\end{equation}
The function $f_s(\xi)$ has asymptotic $\xi^{s}$ at $\xi \to 0$ and a finite
non-zero value at $\xi = \pi/2$:
\begin{equation}
  \label{eq:fs-value-xi-pi2}
  \lim_{\xi \to \pi/2} f_{s}(\xi)  = \frac {\Gamma(s+3/2) \Gamma(1) }{\Gamma(s/2+3/2)\Gamma(s/2+1)}.
\end{equation}

This confirms our argument that there are no non-trivial solutions to the
Laplace equation on $H^3$ with zero asymptotic at the boundary.

Now, given that $\Phi_\R$ and $\Phi_9$ vanish, the linearized equations
(\ref{eq:linearizedA})(\ref{eq:linearizedPhi}) turn into
\begin{gather}
\label{eq:linearized-dA}
  d A = 0 \\
\label{eq:linearized-dPhi}
  d \Phi = 0.
\end{gather}
That means that the complexified gauge connection $A_\BC = A + i \Phi$ is flat.
The third equation in (\ref{eq:linearized-equations}) is effectively a partial
gauge fixing condition on the imaginary part of $A_\BC$. It is actually possible
to rewrite this partial gauge fixing condition in terms of the $d^{*}$ operator
with respect to a rescaled metric. Namely, for the conformally flat
 metric on $\BR^3$ of the form
\begin{equation}
  \label{eq:metric-rescaled}
  g_{ij} = f(|x|)\delta_{ij}
\end{equation}
the $d^{*}_f$ operator acts on one-form $\Phit$ as
\begin{equation}
\label{eq:divergence-metric-f}
   d^{*}_f \Phit = f^{-1} (\p_i \Phit_i + \frac 1 2 f^{-1} f'\Phit_i x_i/|x|),
\end{equation}
where $f'=df(|x|)/dx$. Comparing (\ref{eq:divergence-metric-f}) with
(\ref{eq:linearized-equations3}) we get the scale factor
\begin{equation}
  \label{eq:scale-factored-}
  f(|x|) = \frac {(1+x^2)^2}{(1-x^2)^4}.
\end{equation}
Hence, the partial gauge fixing equation (\ref{eq:linearized-equations3}) is
rewritten as
\begin{equation}
  \label{eq:gauge-fixing-equation-rewritten}
  d^{*}_{f} \tilde \Phi = 0.
\end{equation}

Now we can find all solutions to the linearized problem as follows. From
(\ref{eq:linearized-dPhi}) we solve for $\Phit$ in terms of a scalar potential $p$
\begin{equation}
  \label{eq:Phi-in-terms-potential}
  \Phit = d p.
\end{equation}
The gauge fixing equation (\ref{eq:gauge-fixing-equation-rewritten}) implies then
\begin{equation}
  \label{eq:Laplacian-equation-onP}
  d^{*}_{f} d p  = 0,
\end{equation}
i.e. that $p$ is a harmonic function with respect to the metric
(\ref{eq:scale-factored-}). We can find explicitly the harmonic modes in
spherical coordinates. The metric (\ref{eq:metric-rescaled}) is
\begin{equation}
  \label{eq:f-metric-spherical-coordinates}
  ds^2 = \frac{ d\xi^2 + \sin^2 \xi d\Omega_2^2}{\cos^4 \xi},
\end{equation}
so the Laplacian equation (\ref{eq:Laplacian-equation-onP}) on spherical mode
$p_{s}(\xi)$ with angular momentum $s$ is
\begin{equation}
  \label{eq:Laplacian-equation-on-P-in-spherical-coordinates}
  \cot^2 \xi  \frac{\p} {\p \xi} \left( \tan^2 \xi \frac {\p p_s(\xi)}{\p \xi} \right) -  \frac {s(s+1)}{\sin^2 \xi} p_s(\xi) =  0.
\end{equation}

Again, this is the Laplacian equation in the $(p,q)$ polyspherical coordinates
(\ref{eq:Laplace-equation-polyspherical-coordinates})
with $p=-2,q=2,r=0,l=0$. The solution regular at $\xi=0$ is
\begin{equation}
  \label{eq:solution-for-P-hypergeometric-function}
\begin{split}
  p_{s}(\xi) = \tan^{s} \xi F(s/2, s/2 + 3/2, s+3/2, -\tan^2 \xi) = \\
  =\sin^{s} \xi F(s/2, s/2, s+3/2, \sin^2 \xi).
\end{split}
\end{equation}
The solution is finite at $\xi=\pi / 2$ for any $s$, hence the components of $\Phit$ tangent to
the boundary $\p B^3$ are also finite. To find asymptotic of the normal component of $\Phit$ we
need to know expansion of (\ref{eq:solution-for-P-hypergeometric-function}) at
$\theta = \pi/2-\xi$ at $\theta = 0$. For this purpose we rewrite
(\ref{eq:solution-for-P-hypergeometric-function}) using identity on
hypergeometric functions (see e.g. \cite{Wang-Guo} p.160)
\begin{equation}
  \label{eq:identity-hyper-geometric-functions}
  \begin{split}
  F(\alpha,\beta,\gamma,z) = \frac {
    \Gamma(\gamma)\Gamma(\gamma-\alpha-\beta)}{\Gamma(\gamma-\alpha)
    \Gamma(\gamma-\beta) } F(\alpha,\beta, \alpha+\beta - \gamma+1, 1- z)
 \\ + \frac { \Gamma(\gamma) \Gamma(\alpha+ \beta - \gamma) } {\Gamma(\alpha)
  \Gamma(\beta)} (1-z)^{\gamma-\alpha-\beta} F(\gamma-\alpha, \gamma-\beta,
\gamma-\alpha-\beta+1, 1-z).
  \end{split}
\end{equation}
We get
\begin{equation}
  \label{eq:potential-p-s-near-xi=pi-over-2}
\begin{split}
  p_s(\xi)=\sin^{s}(\xi) \left (
\frac {\Gamma(s+3/2)\Gamma(3/2)}{\Gamma(s/2+3/2)^2} F(s/2,s/2,-1/2,\cos^2 \xi) \right.
\\ + \left.
\frac {\Gamma(s+3/2)\Gamma(-3/2)}{\Gamma(s/2)^2} (\cos^2\xi)^{3/2} F(s/2 + 3/2,
s/2 + 3/2, 5/2 , \cos^2 \xi)
 \right).
\end{split}
\end{equation}
Near $\theta = 0$ we obtain
\begin{equation}
  \label{eq:expansion-of-ps-theta=0}
  p_s(\theta) = \cos^{s} \theta ( A+B \sin^2 \theta + C \sin^3 \theta + \dots),
\end{equation}
where $A,B,C$ some constants. Therefore
\begin{equation}
  \label{eq:derivative-ps-xi}
 \Phit_{\theta} = \frac {\p p_s(\theta)}{\p \theta} = (-As +B) \theta + O(\theta^2).
\end{equation}
This means that the normal component of $\Phit_{}$ at the boundary vanishes as
the first power of $\theta$ or $(1-x^2)$.  Hence, the original scalars, related
to $\Phit$ by (\ref{eq:change-of-variables-PhiI4}), are all finite at the
boundary $S^2$.

So all solutions of the linearized equations
(\ref{eq:linearized-equations})(\ref{eq:linearized-equations2})(\ref{eq:linearized-equations3})
modulo gauge transformations are parametrized by the scalar potential $p$
(modulo zero modes of $p$), which
is a harmonic function in the three-dimensional ball with respect to the metric
(\ref{eq:f-metric-spherical-coordinates}). A harmonic functions $p$ is
uniquely defined by its boundary value on the $S^2$. Hence we see that
that tangent space $T\CalM_0$ to the moduli space of solutions at the origin
is isomorphic to the space of adjoint-valued scalar functions on the $S^2$ modulo zero modes.

\subsubsection{Solution of non-abelian equations: complexified flat connections}

Now we consider the full non-abelian equations
(\ref{eq:my-first-equation-Phi-tilde})(\ref{eq:my-second-equation-Phi-tilde})(\ref{eq:my-third-equation-Phi-tilde}). Looking
back at our solution of the linearized problem (\ref{eq:solution-Phi5-Phi9=0}),
we shall suggest an ansatz $\Phi_R=\Phi_9=0$ for the exact solution.
Then the remaining equations on the complexified connection $A_\BC = A + i \Phit$
are
\begin{gather}
  \label{eq:equations-A-Phi-with-Phi5-Phi9-vanishing}
  F_A - \Phit \wedge \Phit = 0 \\
  d_A \Phit = 0 \\
  d_A^{*_f} \Phit = 0,
\end{gather}
which can be combined into the complexified flat curvature equation
\begin{equation}
  \label{eq:complex-curvature-vanishes}
  F(A_\BC) = 0
\end{equation}
and a partial gauge-fixing equation using the metric
 (\ref{eq:f-metric-spherical-coordinates})
\begin{equation}
  \label{eq:partial-gauge-fixing}
  d_{A} *_{f}  \Phit = 0.
\end{equation}
The first equation can be solved in terms of a scalar function 
$g_\BC: B^3 \to G_{\BC}$, which takes value in the complexified gauge
group  $G_{\BC}$:
\begin{equation}
  \label{eq:solution-in-terms-of-g_c}
  A_\BC = g_\BC^{-1} d g_\BC.
\end{equation}
The partial gauge-fixing condition can be complemented by a real gauge fixing
$d^{*} A = 0$. That gives a non-linear analogue of the harmonic equation (\ref{eq:Laplacian-equation-onP})
\begin{equation}
  \label{eq:partial-complex-gauge-fixing}
  d_A*_{f} (g_\BC^{-1} dg_{\BC}) = 0.
\end{equation}
The solutions of this second order differential equation are parameterized by the
boundary value of $g_{\BC}$. Hence, the tangent space of solutions to the full
non-abelian equations constrained by $\Phi_\R=\Phi_9=0$ coincides with the
moduli space of the linearized problem.

\newcommand{\ggauge}{\mathfrak{g}_{gauge}}
\newcommand{\Ggauge}{G_{gauge}}

We conclude, that the solutions of
(\ref{eq:partial-complex-gauge-fixing}) represent completely moduli space $\CalM$ of
smooth solutions of the supersymmetry equations
(\ref{eq:final-top-equations-original-variables}) with finite action. Hence, the space of gauge
orbits $\M/\Ggauge$ can be parameterized by the boundary value of the
$G_{\BC}/G$-valued potential function $g_\BC$.

Equivalently, we can parameterise $\CalM/\Ggauge$ by the space of complex flat connections on $\Sigma$ modulo the gauge transformations restricted on $\Sigma$
\begin{equation}
  \label{eq:parametrization-boundary}
 \{ A^{2d}_{\BC}| F_{A_{\BC}}  =  0\}. 
\end{equation}

Hence, the localization of the path integral of the four-dimensional $\CalN=4$
SYM theory to the moduli space $\CalM/\Ggauge$ can be represented by
a path integral over the space of complex flat connections on
the $B^3$ boundary $S^2$. The action of this two-dimensional theory
is determined by evaluating  the four-dimensional Yang-Mills functional on the
field configurations representing  $\CalM$.

We will show below that the $\CalN=4$ Yang-Mill action $S_{YM}$ restricted to the 
supersymmetric field configurations is  a total derivative on $B^3$, 
hence it can be expressed in terms of a two-dimensional
action on the boundary $\Sigma$.

We conclude that the outcome of the localization procedure  is a two-dimensional path integral over
the space of complex flat connections on $\Sigma$.

Now we will find the two-dimensional action $S_{2d}$. The measure of integration
in the two-dimensional theory is then $\exp(-S_{2d})$ times the induced volume
form from the four-dimensional theory on the moduli space $\M$.

\section{Two-dimensional theory\label{se:two-dimensional}}

\subsection{The physical action on the supersymmetric configurations}

\subsubsection{The physical action on $B^3 \times_{\w} S^1$  }

The bosonic part of the $\CalN=4$ Yang-Mills action on $S^4$ in
coordinates (\ref{eq:S^4-metric-B3-x-coordinates}) is\footnote{
In all expressions for the action functionals below we do not explicitly write Lie 
algebra indices and the contractions of them by an invariant Killing form $\la , \ra$ on the Lie algebra (it exists
uniquely up to an overall rescaling) but, of course, that is implicitly assumed.
A pedantical reader might wish to substitute 
\begin{equation}
  \label{eq:subst}
 \frac 1 {4 g_{YM}^2} \int \sqrt{g} d^n x  F_{\mu \nu} F^{\mu \nu} \mapsto 
 \frac 1 {4 g_{YM}^2} \int \sqrt{g} d^n x  F_{\mu \nu}^a F^{\mu \nu}_a,  
\end{equation}
where $a,b$ are the Lie algebra indices in an orthogonal basis, e.g. $F = F^{a} T_{a}$ where
$T^a$ are generators of the Lie algebra. For the $\SU(N)$ gauge group
the conventional choice is such that $\tr_F  T_a T_b = -\frac 1 2 \delta_{ab}$.
}

\begin{equation}
  \label{eq:bosonic-YM-action-S4}
\begin{split}
  S_{YM} = \frac {1} {2 g_{YM}^2} \int_{0}^{2\pi} d\tau \int_{|x| <
    1} d^3 x \sqrt{g}
(\frac 1 2 F_{\mu \nu}F^{\mu \nu} + D_{\mu} \Phi_A D^{\mu} \Phi_A  \\ + \frac 1 2
[\Phi_A, \Phi_B]^2 + \frac R 6 \Phi_A^2 + K^2).
\end{split}
\end{equation}
Here $R$ denotes the scalar curvature, which for $S^4$ of radius $1/2$ has value $R =
12/(1/2)^2 = 48$.
First we make Weyl transformation and get the physical action on the space
$B^3 \times_{\w} S^1$ with the metric
(\ref{eq:metric-R3-tau})
\begin{gather}
  \label{eq:Weyl-transformation-D3-to-B3}
  g_{\mu \nu} [S^4] = e^{2\Omega} g[\BR^3 \times_{\w} S^1] \\
  \Phi_A[S^4] = e^{-\Omega} \Phi_A[\BR^3 \times_{\w} S^1] \\
  K_I[S^4] = e^{-2\Omega} K_I[\BR^3 \times_{\w} S^1]
\end{gather}
where
\begin{equation}
  \label{eq:scale-factor-Omega}
  e^{2\Omega} = (1+x^2)^{-2}.
\end{equation}

In terms of the fields on $\BR^3 \times_{\w} S^1$ the bosonic action is
\myref{2008-05-22 p4}
\begin{equation}
\label{eq:action-on-R3}
\begin{split}
  S_{YM} =
\frac 1 {2 g_{YM}^2} \int_{0}^{2 \pi} d \tau \int_{|x|< 1} d^3
x \left(  \frac 1 2 (1-x^2) \times \right. \\
( \frac 1 2 F_{ij}^2 +  g^{\tau \tau} F_{\tau i}^2
+g^{\tau \tau} (D_{\tau} \Phi_A)^2 + (D_{i} \Phi_A)^2 + \frac{2}{(1-x^2)}
\Phi_A ^2 + \frac 1 2 [\Phi_A \Phi_B]^2 + K^2) \\
\left . +  D_i \left( \frac{1-x^2}{1+x^2} x_i \Phi_A^2 \right) \right)
\end{split}
\end{equation}
The last term is the total derivative which vanishes because the factor
$(1-x^2)$ vanishes at the integration boundary $|x|=1$.
The action on  $\BR^3 \times_{\w} S^1$ can be also written starting from
(\ref{eq:bosonic-YM-action-S4})
and substituting the metric (\ref{eq:metric-R3-tau}).
The scalar curvature on $\BR^3 \times_{\w} S^1$ can be computed easily using a general
formula for the scalar curvature on a warped product of two manifold
$M\times_{f} N$, see e.g. \cite{MR896013}.  If $g_{M}$ and $g_{N}$ are the metrics on $M$ and $N$, and
if $g_{M} \oplus f^2 g_{N}$ is the metric on $M \times_{f} N$, then
\begin{equation}
\begin{split}
 R_{M \times_{f} N} u =   -\frac {4n} {n+1} \Delta_M u   + R_M u + R_N
 u^{\frac{n-3}{n+1}}  \\
\text{where} \quad n=\dim N, \quad u = f^{\frac{n+1}{2}}, \quad \Delta_{M}
\, \text{is Laplacian on M}.
\end{split}
\end{equation}

In the case $\BR^3 \times_{\w} S^1$ we get $n = \dim N = 1$, so $u = f = \frac 1 2
(1-x^2)$. Then, for the radius $1/2$, we get
\begin{equation}
R[\BR^3 \times_{\w} S^1] = - u^{-1} \Delta u = \frac {12} {1-x^2},
\end{equation}
which agrees with (\ref{eq:bosonic-YM-action-S4}) and (\ref{eq:action-on-R3}).

Next we rewrite the action in terms of the twisted scalars $\Phi_\T,\Phi_\R$ and
$\Phi_{m}$, $m=6,7,8,9$, \myref{2008-05-22 p5}
(\ref{eq:change-of-variables-for-scalars-Phi5-Phi0})
\begin{equation}
  \label{eq:S_YM-changed-to-PhiT-PhiR}
\begin{split}
  S_{YM} = \frac 1 {2 g_{YM}^2} \int_{0}^{2 \pi} d \tau \int_{|x|< 1} d^3 x
\frac 1 2 (1-x^2)
(  g^{\tau \tau} F_{i \tau}^2 + (D_i \Phi_\T)^2  \\
+  g^{\tau \tau} (D_{\tau} \Phi_\R - \Phi_\T)^2 + [\Phi_\T,\Phi_\R]^2 +
   g^{\tau \tau} (D_{\tau} \Phi_{m})^2 + [\Phi_\T, \Phi_{m}]^2  \\
   g^{\tau \tau} (D_{\tau} \Phi_\T + \Phi_\R )^2 + \frac 1 2 F_{ij}^2 + (D_{i}
   \Phi_{m})^2  + (D_{i} \Phi_\R)^2  + \frac 1 2 [\Phi_m, \Phi_n]^2 +
   [\Phi_\R ,\Phi_m]^2 \\  + \frac{2}{(1-x^2)}
(\Phi_m^2 + \Phi_\T^2 + \Phi_\R^2) + K_I^2 ).
 \end{split}
\end{equation}

\subsubsection{The physical action reduced to the $B^3$ }

Then we restrict the action onto configurations invariant under 
the diagonal $\U(1)_S \subset \SO(2)_S \times  \SO(2)_B$ 
 using (\ref{eq:summmary-QPsi-bottom}) and
(\ref{eq:QPsi-top-tau-T}). We also assume that $\Phi_\T = 0$ in the
supersymmetric background, otherwise $\Phi_\T$ has first order singularity near
the $S^2$ which would mean insertion of surface operator.
Removing the terms with $\nabla_{\tau}$ and $\Phi_\T$ from the action
(\ref{eq:S_YM-changed-to-PhiT-PhiR}),
we arrive to this three-dimensional action for the gauge field $A_{i}$ and five
scalars $\Phi_{\R}, \Phi_m$, $m=6,7,8,9$,
\begin{equation}
  \label{eq:S_YM-on-B3}
\begin{split}
  S_{YM}^{\text{inv}}(B^3) = \frac 1 {2 g_{YM}^2} 2 \pi \int_{|x|< 1} d^3 x
\frac 1 2 (1-x^2)
(   \frac{4}{(1-x^2)^2}  \Phi_\R^2 + \frac 1 2 F_{ij}^2 + (D_{i} \Phi_{m})^2 \\ + (D_{i} \Phi_\R)^2  + \frac 1 2 [\Phi_m, \Phi_n]^2 +
   [\Phi_\R ,\Phi_m]^2  + \frac{2}{(1-x^2)}
(\Phi_m^2  + \Phi_\R^2) + K_I^2 ).
 \end{split}
\end{equation}

\subsubsection{The boundary term}

Now we show that modulo the supersymmetry equations the physical action (\ref{eq:S_YM-on-B3})
 on the $U(1)_S$ invariant configurations reduced to  $B^3$  is a total derivative.
We try the following ansatz
\begin{multline}
\label{eq:S-susy-B3}
  S_{susy}^{\text{inv}}(B^3) = \frac {1} {4 g_{YM}^2} 2 \pi \int_{|x|<1} d^3 x  \\
(
(-\frac 1 2 (F_{ij} - [\Phi_{i+4} \Phi_{j+4}]) \ep_{ijk} + K_5 x_k - K_{i+4} x_j \ep_{ijk}) \times \\
(-\frac 1 2 (F_{ij} - [\Phi_{i+4} \Phi_{j+4}]) \ep_{ijk} - K_5 x_k + K_{i+4} x_j \ep_{ijk})
\\+
(\nabla_{i} \Phi_{i+4} - K_i x_i)(\nabla_j \Phi_{j+4} + K_j x_j)
\\+
((\nabla_i \Phi_{j+4} - K_i x_j) \ep_{ijk} )(\delta_{k \tilde k} - x_{k} x_{\tilde k})
( (\nabla_{\tilde i} \Phi_{\tilde j+4} + K_{\tilde i} x_{\tilde j}) \ep_{\tilde i \tilde j \tilde k})
\\+
(K_k - (x_i \nabla_i \Phi_{k+4} + x_i \nabla_k \Phi_{i+4} - x_k \nabla_{i} \Phi_{i+4} + 2 \Phi_{k+4})) \times \\
(K_k + (x_i \nabla_i \Phi_{k+4} + x_i \nabla_k \Phi_{i+4} - x_k \nabla_{i} \Phi_{i+4} + 2 \Phi_{k+4}))
\\+
(K_5 + \frac 1 2 x_k \ep_{ijk}(F_{ij} - [\Phi_{i+4} \Phi_{j+4}]))
(K_5 - \frac 1 2 x_k \ep_{ijk}(F_{ij} - [\Phi_{i+4} \Phi_{j+4}]))
\\+
(K_{k+4} + x_i(F_{ik} + [\Phi_{i+4} \Phi_{k+4}]))(K_{k+4} - x_i (F_{ik} + [\Phi_{i+4} \Phi_{k+4}]))
\\
-(x_{i} K_{i+4})^2.
)
\end{multline}
Each term above corresponds to one of the top
supersymmetry equations (\ref{eq:extraK}),(\ref{eq:K-fields-in-terms-of-APHI}) and (\ref{eq:APHI-with-K-top-equations})
multiplied by a suitable factor to match the kinetic term of the reduced Yang-Mills action (\ref{eq:S_YM-on-B3}).
Therefore at the supersymmetric configurations $S_{susy}^{\text{inv}}(B^3)$ vanishes.
After some algebra\myref{2008-06-21 p8}, one can show that the actions (\ref{eq:S_YM-on-B3})
and (\ref{eq:S-susy-B3}) differ on a total derivative
\begin{multline}
\label{eq:S-YM-difference-S-susy}
 S_{susy}^{\text{inv}}(B^3) = S_{YM}^{\text{inv}}(B^3) + \frac {2\pi} {4 g^2_{YM}} \int {d^3 x}_{|x|<1}
( \nabla_i ((1-x^2) \Phi_{i+4} \nabla_{j} \Phi_{j+4} - \Phi_{j+4} \nabla_{j} \Phi_{i+4}) \\
- 4 \nabla_{j}( x_i x_k \Phi_{k+4} \nabla_{i} \Phi_{j+4} - x_i x_j \Phi_i \nabla_{k+4} \Phi_{k+4}) \\
- 6 \nabla_{j}(x_i \Phi_{i+4} \Phi_{j+4}))
\end{multline}
Integrating the total derivative term we get a boundary action
\begin{multline}
\label{eq:S-YM-S-S-susy-difference-boundary-action}
  S_{YM}^{\text{inv}}(B^3) = S_{susy}^{\text{inv}}(B^3) +  \frac {2\pi} {4 g^2_{YM}} \int_{S^2:\, |x|=1} d\Omega \,(4 \Phi_n ( \nabla_{n} \Phi_n - \nabla_{i} \Phi_{i+4}) + 6 \Phi_n^2),
\end{multline}
where $\Phi_n$ is the normal component to the $S^2$ of the one-form $\Phi$, i.e. $\Phi_{n} = n_i \Phi_{i+4}$,
and $\nabla_n$ is the derivative in the normal direction, $n_i = x_i /|x|$.
Using the equation (\ref{eq:APHI-with-K-top-equations}) for $\Re Q\Psi^{t}|_{e_5}$ with $K_i$ substituted from (\ref{eq:K-fields-in-terms-of-APHI})
we get a constraint on $\Phi_n$ on the boundary
\begin{equation}
\label{eq:Phi-n-nabla-Phi-relation}
  \nabla_{n} \Phi_{n} - \nabla_{i} \Phi_{i+4} = -\Phi_{n}.
\end{equation}
Hence, the boundary action (\ref{eq:S-YM-S-S-susy-difference-boundary-action}) simplifies to
\begin{equation}
  S_{YM}^{\text{inv}}(B^3) = S_{susy}^{\text{inv}}(B^3) + \frac {\pi} {g^2_{YM}} \int_{S^2: |x|=1} d\Omega \,  \Phi_n^2,
\end{equation}
where $d\Omega$ is the standard volume form on $S^2$.
On supersymmetric configuration $S_{susy}^{\text{inv}}(B^3)$ vanishes, thus the $\CalN=4$ Yang-Mills localizes to the two-dimensional theory
on $S^2$ with the action
\begin{equation}
  S_{2d} = \frac {\pi}{g^2_{YM}} \int_{S^2: |x|=1} d\Omega \, \Phi_n^2.
\end{equation}
Equivalently we can express the action in terms of the tangent to $S^2$ components of $\Phi$ using the
 constraint (\ref{eq:Phi-n-nabla-Phi-relation})
\begin{equation}
  \label{eq:S-2d-equiv}
  S_{2d} = \frac {\pi}{g^2_{YM}} \int_{S^2: |x|=1} d\Omega \, (d_A^{*2d} \Phi_{t})^2,
\end{equation}
where $\Phi_t$ denotes an adjoined-valued one-form on $\Sigma$ obtained from the components of $\Phi_{i}$ tangential 
to $\Sigma$. To get (\ref{eq:S-2d-equiv}) we used (\ref{eq:Phi-n-nabla-Phi-relation}) and
the relation between the tangential components of $\Phi$ in $\BR^3$ coordinates with the one-form $\Phi_{t}$
\begin{equation}
  \label{eq:curved-flat}
  \nabla_{i} \Phi_{i+4} - \nabla_{n} \Phi_{n} = d^{*2d}_A \Phi_{t} + 2 \Phi_{n},
\end{equation}
from which one gets that
\begin{equation}
  \label{eq:f-Phin-relation}
  d^{*2d}_{A} \Phi_{t} = - \Phi_n \quad \text{on supersymmetric configurations}.
\end{equation}

We recall that the scalar fields in (\ref{eq:Weyl-transformation-D3-to-B3}) - (\ref{eq:S-2d-equiv}) are the fields for the four-dimensional
theory on $\BR^3 \times_{\w} S^1$. In terms of the original fields of the $\CalN=4$ Yang-Mills on $S^4$ we have $\Phi[\BR^3 \times_{\w} S^1]=(1+x^2)^{-1}
\Phi[S^4]$, so
\begin{equation}
  S_{2d} = \frac {\pi}{4 g^2_{YM}} \int_{S^2: |x|=1} d\Omega (d_A^{*2d} \Phi_{t}^{S^{4}})^2.
\end{equation}
Above was assumed that the radius $r=\frac 1 2$. To restore $r$ we need to insert a power of factor $(2r)$ to get the correct dimension
\begin{equation}
\label{eq:two-dimensional-action}
  S_{2d} = (2r)^2 \frac {\pi}{4 g^2_{YM}} \int_{S^2: |x|=2r} \sqrt{g_{S^2}} d^2 \sigma (d_A^{*2d} \Phi_{t}^{S^{4}})^2.
\end{equation}

\subsubsection{Relation to the constrained 2d complexified Yang-Mills }

In this section $\Phi$ denotes the one-form on $\Sigma$ previously called $\Phi_{t}$. 
The Wilson loop operator (\ref{eq:Wilson-loops-S3}) descends to the Wilson loop operator in the two-dimensional theory
\begin{equation}
  \label{eq:Wilson-loop-2d}
  W_R(C) = tr_{R} \Pexp \oint  (A - i * \Phi)
\end{equation}

\newcommand{\Ac}{\tilde A_\BC}
We introduce another complexified connection 
\begin{equation}
  \label{eq:complexified-connection}
  \Ac = A - i * \Phi,
\end{equation}
so the Wilson loop operator  (\ref{eq:Wilson-loop-2d}) is the holonomy of $\Ac$
\begin{equation}
  \label{eq:Wilson-loop-2d}
  W_R(C) = \tr_{R} \Pexp \oint \Ac.
\end{equation}

Let $F_{\Ac}$ be the curvature of $\Ac$, then
\begin{equation}
  F_{\Ac} = d \Ac + \Ac \wedge \Ac = F_A - \Phi \wedge \Phi - id_A * \Phi.
\end{equation}
By (\ref{eq:equations-A-Phi-with-Phi5-Phi9-vanishing}) at the localized configurations we have $F_A - \Phi \wedge \Phi = 0$, then
\begin{equation}
  d_A * \Phi = i F_{\Ac} \quad \text{for localized configurations}.
\end{equation}

Then the action of the two-dimensional theory (\ref{eq:two-dimensional-action}) is equivalent to the action of the bosonic Yang-Mills for complexified connection $\Ac$
\begin{equation}
\label{eq:two-dimensional-action-Ac}
  S_{2d} =  + \frac {1} {2 g_{2d}^2} \int_{S^2} d\Omega \,( *_{2d} F_{\Ac})^2,
\end{equation}
where the two-dimensional coupling constant is denoted $g_{2d}$
\begin{equation}
  \label{eq:g-YM-g-2d-relation}
  g_{2d}^2 = -\frac {g_{YM}^2} {2 \pi r^2}.
\end{equation}
This relation agrees with the conjecture \cite{Drukker:2007qr,Drukker:2007dw,Drukker:2007yx} 
given that $g_{2d}^2$ is properly defined in the 2d YM action.\footnote{We write the action in terms of the scalar field $*_{2d} F$ which is the Hodge dual to the curvature
two form $F$. 
In components one has $(*_{2d} F)^2  = \frac 1 2 F_{\mu \nu} F^{\mu \nu}$ which means that we use the same 
conventions for the normalization of the 2d YM action as for the 4d YM action (\ref{eq:bosonic-YM-action-S4}), i.e.
$S = \frac {1} {4 g^2} \int d^n x \sqrt{g} F_{\mu \nu} F^{\mu \nu} = 
-\frac 1 {2 g^2} \int d^n x \sqrt{g} \tr F_{\mu \nu} F^{\mu \nu}$ for $\SU(N)$ gauge group}

So the original four-dimensional problem has been reduced to complexified two-dimensional bosonic Yang-Mills theory (\ref{eq:two-dimensional-action-Ac})
with the standard Wilson loop observables (\ref{eq:Wilson-loop-2d}).  The
 complexified connection $\Ac = A -i * \Phi$ is constrained by (\ref{eq:equations-A-Phi-with-Phi5-Phi9-vanishing})
\begin{gather}
\label{eq:two-dimensional-constraints}
\Re F_{\Ac} = 0 \\
d_{\Re \Ac} * \Im \Ac  = 0.
\end{gather}
The two real constraints remove two real degrees of freedom from
the four real degrees of freedom of complex one-form $\Ac$ (we do not subtract gauge symmetry in this counting).
Therefore, the path integral is taken over a certain half-dimensional subspace of complexified connections $\Ac$.

We can interpret the path integral for the usual two-dimensional Yang-Mills for
real connections as a contour integral in the space of complexified connections,
where the contour is given by the constraint that the imaginary part of the
connection vanishes: $\Im \Ac = 0$.

Our assertion is that the complexified theory
(\ref{eq:two-dimensional-action-Ac}) with constraints
(\ref{eq:two-dimensional-constraints}) is equivalent to the real theory by a
change of the integration contour in the space of complexified connections.

Since perturbative correlation functions of holomorphic observables do not
depend on deformation of the contour of integration, we conclude that the
expectation value of Wilson loop observables (\ref{eq:Wilson-loop-2d})
perturbatively coincides with the expectation values of Wilson loops in
the ordinary two-dimensional Yang-Mills.

We shall look at the constrained complexified two-dimensional Yang-Mills theory
from slightly broader viewpoint of so called topological Higgs-Yang-Mills theory
\cite{Moore:1997dj,Gerasimov:2006zt,Gerasimov:2007ap} which deals with the moduli space of solutions to Hitchin equations. \myref{2008-06-03}

\subsection{Hitchin/Higgs-Yang-Mills theory}

Here we will review Hitchin/Higgs-Yang-Mills theory~\cite{MR887284,MR885778} following
\cite{Moore:1997dj,Gerasimov:2006zt,Gerasimov:2007ap,Kapustin:2006pk}.
Let $\Sigma$ be a Riemann surface, $A$ be a gauge field for the gauge group $G$
($G$ is a compact Lie group) and $\Phi$ be a one-form taking value in the Lie algebra $\g$ of $G$.

Let $\varphi$ be a scalar field taking value in $\g$. The field $\varphi$ can be thought as
an element of the Lie algebra $\ggauge$ of the infinite-dimensional  group of gauge
transformations $\Ggauge$. Let $M$ be the space of fields $(A,\Phi)$.
Using the invariant Killing form on $\g$ we identify $\g$ with $\g^{*}$. Then
locally $M$ is $T^* \Omega^{1} (\Sigma, \ad \g)$.

We notice (see~\cite{Witten:1992xu,Gerasimov:2006zt,Gerasimov:2007ap,Moore:1997dj,Kapustin:2006pk}) that
the space $M$ can be equipped with a triplet of symplectic
structures $\omega_{i}$ and a triplet of corresponding Hamiltonian moment maps
$\mu_{i}$ for $\Ggauge$ acting on $M$.

Explicitly we define the symplectic structure $\omega_{i}$ as follows. Let
$\delta$ be the differential on $M$. Then\footnote{Here the subscripts $1,2$ enumerate arguments of the functional two-form $\omega$, but not the coordinates on $\Sigma$.}
\begin{align}
  \label{symplectic-structures-triplet1}
  \omega_1(\delta A_1, \delta \Phi_1; \delta A_2, \delta \Phi_2)& =  \int_{\Sigma}
  \delta A_1 \wedge \delta A_2 - \delta \Phi_1 \wedge \delta \Phi_2  \\
  \label{symplectic-structures-triplet2}
  \omega_2(\delta A_1, \delta \Phi_1; \delta A_2, \delta \Phi_2)& =  \int_{\Sigma}
  \delta A_1 \wedge \delta \Phi_2 - \delta A_2 \wedge \delta \Phi_1  \\
  \label{symplectic-structures-triplet3}
  \omega_3(\delta A_1, \delta \Phi_1; \delta A_2, \delta \Phi_2)& =  \int_{\Sigma}
  \delta A_1 \wedge *\delta \Phi_2 - \delta A_2 \wedge *\delta \Phi_1,
\end{align}
where $*$ is the Hodge star on $\Sigma$. 

A functional $\mu: M \to \ggauge^*$ is called a moment map if
\begin{equation}
  \label{eq:moment-map-definition}
  i_{\phi} \omega = \mu(\phi) \quad \text{for all} \quad \phi \in \ggauge,
\end{equation}
where $i_{\phi}$ denotes a contraction with a vector field generated on $M$ by an
element $\phi \in \ggauge$.

The group $\Ggauge$ acts on $M$ by the usual gauge transformations
\begin{equation}
  \label{eq:gauge-transformation-A-Phi}
  \begin{aligned}
  \delta A = -d_{A} \phi \\
  \delta \Phi  = [\phi, \Phi].
  \end{aligned}
\end{equation}

One can check  that the functionals
\begin{align}
\label{eq:moment-maps}
  \mu_{1}(\phi)& =  \int (\phi, F - \Phi \wedge \Phi) \\
  \mu_{2}(\phi)& =  \int (\phi, d_A \Phi) \\
  \mu_{3}(\phi)& =  \int (\phi, d_A *\Phi)
\end{align}
are the moment maps for the symplectic structure $\omega_{1}, \omega_{2},
\omega_{3}$ correspondingly.

The space $M$ has natural linear flat structure and the corresponding flat
metric is
\begin{equation}
  \label{eq:flat-metric-on-M-A-Phi}
  g(\delta A_1, \delta \Phi_1; \delta A_2, \delta \Phi_2) =  \int \delta A_1
  \wedge * \delta A_2 + \delta \Phi_1 \wedge * \delta \Phi_2.
\end{equation}

Using the metric $g$ on $M$, to each symplectic structure $\omega_i$ we can
associate a complex structure $I_i$ in the usual way $\omega(\cdot, \cdot) = g(I
\cdot, \cdot)$.

Comparing
\begin{equation}
  \int_{\Sigma} I(\delta A_1) \wedge *\delta A_2 + I(\delta \Phi_1) \wedge
 *\delta \Phi_2
\end{equation}
with
(\ref{symplectic-structures-triplet1})-~(\ref{symplectic-structures-triplet3})
we get
\myref{2008-06-03 p4}
\begin{align}
  I_1(\delta A)& = * \delta A &   I_1(\delta \Phi)& = -* \delta \Phi \\
  I_2(\delta A)& = * \delta \Phi & I_2(\delta \Phi)& = * \delta A \\
  I_3(\delta A)& = - \delta \Phi & I_3(\delta \Phi)& = \delta A
\end{align}

The following linear combinations span the holomorphic subspaces
($+i$-eigenspaces) of the corresponding complex structures:

\begin{equation}
  \label{eq:holomorphic-I1-I2-I3}
  \begin{aligned}
    I_1( A - i*A )   & = i(A - i*A) \\
    I_2( A - i*\Phi) & = i(A - i*\Phi) \\
    I_3( A + i \Phi) & = i( A + i \Phi).
  \end{aligned}
\end{equation}

One can also check that the complex structures satisfy
$I_3 = I_2 I_1, I_1 = I_3 I_2, I_2 = I_1 I_3$. Hence the space $M$ is the
hyperKahler space.

We can use four-dimensional notations. Let us denote
\begin{equation}
  \label{eq:renotation-A-Phi}
  \Phi_1 \equiv A_4 \quad  \Phi_2 \equiv A_3,
\end{equation}
then the three moment maps~(\ref{eq:moment-maps}) correspond to the components
of the self-dual part $F_A^{+}$ of the four-dimensional curvature $F_A$:
\begin{equation}
  \label{eq:moment-maps-4d}
  \begin{aligned}
  F - \Phi \wedge \Phi& = (F_{12} + F_{34}) dx^1 \wedge dx^2 \\
  d_A \Phi & = (F_{13} - F_{24}) dx^1 \wedge dx^2 \\
  d_A *\Phi& = (F_{14} + F_{23}) dx^1 \wedge dx^2
  \end{aligned}
\end{equation}

Clearly, the space $\BR^4$  (or more generally $T^*\Sigma$) is hyperKahler, so it is equipped with $\CP^{1}$ family
of complex structures. Let $z_1,\bar z_1, z_2, \bar z_2$ be complex coordinates
with respect to some complex structure, e.g. $z_1 = x_1 + i x_2, z_2 =
x_3 + i x_4$. Then, in terms of $A_{\bar z_1} = \frac 1 2 (A_{1} + iA_{2})$,
etc, we can write
\begin{equation}
  \label{eq:moment-maps-complex-notations}
  \begin{aligned}
  F_{z_1 \bar{z_1}} + F_{z_2 \bar{z_2}}& = \frac i 2 (F_{12} + F_{34}) =
  \frac i 2  \mu_1  \\
  F_{\bar{z_1} \bar{z_2}} & = \frac 1 4 (F_{13} - F_{24}) + \frac {i} {4}
  (F_{23} + F_{14}) = \frac 1 4 ( \mu_2 + i \mu_3)
  \end{aligned}
\end{equation}

\subsubsection{Constrained Higgs-Yang-Mill theory: cHYM and aYM\label{se:cHYM-aYM}}
For the related story see \cite{Gerasimov:2006zt,Gerasimov:2007ap}.

Consider the following path integral over $\phi$ and the space $M$ of fields $(A,\Phi)$
\begin{equation}
  \label{eq:cHYM-2d}
  Z_{cHYM} = \int_{M|\mu_1 = \mu_2 = 0} D\phi  e^{ i( \omega_3 -
    \mu_3 (\phi))  - \frac {t_2} {2} \int   \phi ^2}.
\end{equation}

The constraints $\mu_1 = \mu_2 = 0$ mean that we set to zero the complexified curvature $F_{A_\BC} = 0$. 
After integrating out $\phi$ one gets the same action as (\ref{eq:two-dimensional-action}). 

In this work we did not compute the one-loop determinant associated with the localization, hence
we do not have a rigorous and complete understanding of the resulting two-dimensional theory. 
However, the most natural assumption is that this determinant in the $\CalN=4$ theory is trivial 
in the same way as in \cite{Pestun:2007rz}. Let us assume that the proper treatment of 
the one-loop determinant and careful consideration of the fermions will lead to the 
constrained Hitchin-Yang-Mills theory (\ref{eq:cHYM-2d}), we call it cHYM theory.

Later we will insert Wilson loop observables  for the holomorphic part of the
complexified connection with respect to the
complex structure $I_2$. Explicitly such observables have form
\begin{equation}
  \label{eq:W-loop-observables-cHYM}
  W_R(C) = \tr_R \Pexp \oint_{C} (A - i * \Phi),
\end{equation}
were $C$ is a contour on $\Sigma$ and $R$ is representation of $G$.

We would like to look at the cHYM theory as a ``hyperKahler rotation'' of another theory
\begin{equation}
  \label{eq:2dYM-cHYM-notations}
  Z_{aYM} = \int_{M| \mu_2 =\mu_3 = 0} D\phi e^{ i(\omega_1 -
    \mu_1(\phi)) - \frac {t_2} {2} \int \phi^2},
\end{equation}
which is almost equivalent  to the bosonic two-dimensional Yang-Mills, hence we
refer to it as aYM theory. Let $\Sigma$
be a Riemann sphere. The constraint $\mu_2 = \mu_3 = 0$ means $d_A^*\Phi = d_A
\Phi = 0$. For a generic connection $A$, the only solution to these constraints
is $\Phi=0$. Then the path integral (\ref{eq:2dYM-cHYM-notations}) reduces to
the 2d bosonic Yang-Mills integral over $A$ and $\phi$ written in the first order formalism
as in \cite{Witten:1992xu}.

We can insert  Wilson loop observables (\ref{eq:W-loop-observables-cHYM}) into the path integral.
Since $\Phi$ vanishes because of the constraint, the Wilson loop
(\ref{eq:W-loop-observables-cHYM}) reduces to
 the ordinary Wilson loop of the connection $A$. Therefore, the expectation value of
 Wilson loops (\ref{eq:W-loop-observables-cHYM}) naively is computed by the standard formulas of the
two-dimensional Yang-Mills theory
\cite{Migdal:1975zf,Witten:1992xu,Witten:1991we} modulo subtleties which are
related to non-generic connections for which there are non-trivial solutions of
the  constraint $d_A^*\Phi = d_A \Phi = 0$. Such connections precisely
correspond to unstable instantons, i.e. configurations with covariantly constant
curvature $F_A$. It is well known that the partition function of bosonic
two-dimensional Yang-Mills can be written as a sum of contributions from
such unstable instantons \cite{Witten:1992xu,Gerasimov:1993ws,Blau:1993hj}.
A contribution of a classical solution with  curvature $F$ enters with
a weight $\exp(-\frac 1 {2 g^2} \rho(\Sigma) F^2 )$ where $\rho(\Sigma)$ is the
area of $\Sigma$. In the weak coupling limit such instanton contributions are
exponentially suppressed and do not contribute to the perturbation theory.
Hence, we conclude that perturbatively the aYM theory
(\ref{eq:2dYM-cHYM-notations}) is equivalent to the ordinary two-dimensional
Yang-Mills. 

However, at the non-perturbative level, the  aYM theory is different from the usual 2d YM theory.
Here we assume the gauge group $G = U(N)$ and consider topologically trivial situation $c_1(E) = 0$ where 
E is the gauge bundle. And we take $\Sigma = S^2 \simeq \CP^1$. 
If $A$ is a connection corresponding to an ``unstable instanton'', the holomorphic 
vector bundle $E$  associated to $A$ splits as a sum of nontrivial line bundles $\CalO(n_1) \oplus \dots \oplus \CalO(n_N)$, for integers $n_1 + \dots + n_N = 0$. 
Then the equation $d_A^*\Phi = d_A \Phi = 0$ has non-trivial solutions for $\Phi$, 
and as well there are non-trivial zero modes for associated fermions.\footnote{We do not write the action for fermions in 
this section but assume that it is the natural as one can find in e.g. 
\cite{Gerasimov:2006zt,Gerasimov:2007ap,Moore:1997dj}.} One can see this by writing the one-form $\Phi$
as $\Phi = \Phi_z dz + \Phi_{\bar z} \overline{dz}$. The two real equations $\mu_2 = \mu_3 = 0$
are equivalent to the one complex equation $ [ \p_{\bar z} + A_{\bar z}, \Phi_z ] = 0$, 
which means that $\Phi_z$ is adjoined-valued holomorphic one-form. The field $\Phi_z$ 
represents a section of $\mathrm{Ad(E)} \otimes T^*_{\Sigma}$ where $T^*_{\Sigma}$ denotes 
the holomorphic cotangent bundle on $\Sigma$. On $\Sigma = \CP^{1}$ one has $T^{*}_{\Sigma} \simeq \CalO(-2)$. 
If the bundle $E$ splits as $\CalO(n_1) \oplus \dots \oplus \CalO(n_N)$ then the $N^2$ dimensional 
bundle $\mathrm{Ad}(E)$ associated with the adjoint representation splits as $\oplus_{i,j = 1\dots N} \CalO(n_i - n_j)$.
A bundle $\CalO(n)$ has holomorphic sections only for non-negative $n$.
Therefore, one concludes that if the connection $A$ is generic, and hence the bundle $E$ is a  holomorphically trivial
bundle with $n_1 = \dots =n_N$, then there are no nontrivial holomorphic sections of $\mathrm{Ad}(E) \otimes T_{\Sigma}^*$, 
and thus $\Phi$ and the associated fermions must vanish. However, if the connection $A$ corresponds
to an ``unstable instanton'', and hence the bundle $E$ is a holomorphically non-trivial bundle
with some of $n_k \neq 0$, then there exist a non-zero holomorphic sections 
of $\mathrm{Ad}(E) \otimes T_{\Sigma}^*$ as well as there are some zero-modes for associated fermions. 
We assume that the path integral of aYM theory can be localized to the 2d instanton connections 
like in the case of the usual 2d YM theory, but that unstable instantons do not actually contribute 
because of the fermionic zero mode which appear for holomorphically non-trivial bundles $E$.

\subsubsection{From cHYM to aYM  perturbatively}
Let us give more details supporting the claim that the perturbative expectation value
 of Wilson loop (\ref{eq:W-loop-observables-cHYM}) in the cHYM theory (\ref{eq:cHYM-2d})
and the aYM theory (\ref{eq:2dYM-cHYM-notations}) is the same.

\paragraph{aYM theory.} First we consider the aYM theory (\ref{eq:2dYM-cHYM-notations}). 
We write the constraints $\mu_2 = \mu_3 = 0$ using Lagrange multipliers.
We introduce scalar auxiliary fields $H_2, H_3$ and their
superpartners $\chi_2, \chi_3$. The superpartners of $A$ and $\Psi$ are
fermionic adjoined valued one-forms on $\Sigma$.
Then we consider the usual complex for equivariant cohomology
\begin{equation}
\begin{aligned}
  Q A &= \psi_A &                &Q \chi_{2,3} = H_{2,3} \\
  Q \psi_A &= -d_A \phi &     &Q H_{2,3} = [\phi,\chi_{2,3}] \\
\end{aligned}
\end{equation}
with
\begin{equation}
  Q \phi = 0.
\end{equation}

The aYM theory (\ref{eq:2dYM-cHYM-notations}) can be rewritten as
\begin{multline}
  \label{eq:2dYM-cHYM-fermionic-notations}
  Z_{aYM} = \int D\phi \, DAD\psi_A \, D\Phi D\psi_\Phi DH D\chi \\
 \exp(
 \int i(\psi_A \wedge \psi_A  - \psi_\Phi \wedge \psi_\Phi - (F-
 \Phi \wedge \Phi))\phi - \frac {t_2} 2  \phi \wedge * \phi \\
 + S_{c}),
\end{multline}
where
\begin{multline}
  \label{eq:S-constraint}
  S_{c} = i Q ( \int d_A \Phi \wedge \chi_2  + d_A  *\Phi \wedge \chi_3) =  \\
  i \int  ( d_A \psi_\Phi + [\psi_A, \Phi]) \wedge  \chi_2 +
        ( d_A *\psi_\Phi+ [\psi_A, *\Phi]) \wedge  \chi_3 +
         d_A \Phi \wedge  H_2 +  d_A *\Phi \wedge H_3.
\end{multline}

If we integrate out the Lagrange multipliers $H_2,H_3$ and $\Phi$, and their
fermionic partners $\chi_2, \chi_3$ and $\psi_A$, the resulting determinants
cancel, while $\Phi$ becomes restricted to the slice $d_A \Phi = d_A^* \Phi =
0$, and similarly $\psi_\Phi$ is restricted to $d_A \psi_\Phi + [\psi_A,\Phi]=0$
and $d_A * \psi_\Phi + [ \psi_A, * \Phi] = 0$. Since $\Phi =0$ we get
$\psi_\Phi=0$. Then what remains is
\begin{equation}
  Z_{aYM} = \int DA D\psi_A D\phi \exp ( \int i( \psi_A \wedge \psi_A - F\phi) -\frac
  {t_2} 2 \phi \wedge *\phi),
\end{equation}
which is the usual action of bosonic Yang-Mills in the first order formalism \cite{Witten:1992xu}.
In this derivation we have been careless in assuming that $d_A \Phi = d_A^* \Phi = 0$ implies $\Phi=0$, 
which is true for a generic connection but not for unstable instantons as discussed in \ref{se:cHYM-aYM}.
Therefore, here we only claim that aYM theory is  equivalent to the YM up to the instanton corrections.

\paragraph{cHYM theory.} Now consider the cHYM theory (\ref{eq:cHYM-2d})\myref{2008-06-13}.
First we write it in a slightly different way:
\begin{equation}
  \label{eq:cHYM-2d-version}
  Z_{cHYM} = \int_{M|\mu_1 = \mu_2 = 0} D\phi  e^{ i( \omega_3 + i\omega_1 -
    (\mu_3 (\phi)+ i \mu_1(\phi))  - \frac {t_2} {2} \int  \phi ^2}.
\end{equation}
Here we added to the action the term $\mu_1(\phi)$ and its supersymmetric extension $\omega_1$.
Since $\mu_1(\phi)=0$ by constraint, classically this is the same theory as (\ref{eq:cHYM-2d}), 
and we assume the proper treatment of fermions makes this claim valid also on quantum level. 
The symplectic structure $\omega_1 - i\omega_3$ is the holomorphic $(2,0)$ two-form
with respect to the second complex structure in (\ref{eq:holomorphic-I1-I2-I3}).

\newcommand{\tA}{\tilde A}
Let us make a change of variables in the path integral from the fields $(A, \Phi)$ to the fields $(\Ac,\Phi)$
where
\begin{equation}
  \label{eq:change-variables}
  \Ac = A - i*\Phi. \\
\end{equation}
Perturbatively we can rotate the integration contour for $\Phi$ to the imaginary
axis, then $\Ac$ is real valued.
The Jacobian for this change of variable is trivial.

The symplectic structure $\omega_1 - i\omega_3$ can be written as
\begin{equation}
  \label{eq:symplectic-structure-13}
  \omega_1 - i\omega_3 =  \int_{\Sigma} \delta \Ac \wedge \delta \Ac,
\end{equation}
and the moment map $\mu_1 - i\mu_3$ is actually the curvature of $\Ac$
\begin{equation}
  \label{eq:mu-1-mu-3-moment-maps}
  \mu_1 -i \mu_3 = F(\Ac)
\end{equation}

One can see that if $\Sigma$ is a sphere, then constraints $\mu_1 = 0, \mu_2=0$
determine $\Phi$ uniquely for each $\Ac$. Hence, the path integral (\ref{eq:cHYM-2d-version}) reduces to
the integral over the fields $\Ac$ with the measure induced by the symplectic
structure (\ref{eq:symplectic-structure-13}). That is the standard bosonic
Yang-Mills theory in the first order formalism for the connection $\Ac$.
The correlation function of Wilson loop operators
(\ref{eq:W-loop-observables-cHYM}) perturbatively are computed as in the usual
bosonic two-dimensional Yang-Mills.

%\section{Conclusion \label{se:Conclusion}}

\subsubsection{Remarks and outlook}

In~\cite{Witten:1992xu} Witten has related the physical two-dimensional Yang-Mills
theory (\ref{eq:2dYM-cHYM-notations}) with the topological two-dimensional
Yang-Mills. The key point is that the path integral for the physical Yang-Mills
theory can be represented as an integral of the equivariantly closed form with
respect to the following operator $Q$
\begin{equation}
  \label{eq:operator-Q-2d-bosonic-Yang-Mills}
  \begin{aligned}
    Q A& = \psi \\
    Q \psi& = -d_A \phi \\
    Q \phi& = 0.
  \end{aligned}
\end{equation}
In other words, the $\omega_1 - \mu_1(\phi)$ is the equivariantly closed form
constructed from the symplectic structure $\omega_1$ and the Hamiltonian moment
map $\mu_1$ for the gauge group acting on the space of connections. Then
localization method can be used to compute the integral of such equivariantly
closed form~\cite{MR721448,MR685019,MR674406,Witten:1992xu}.

Though the Wilson loop observable is not $Q$-closed, its expectation value can
be still solved exactly. That gives a hope that we can also find exact expectation value of Wilson loops
(\ref{eq:W-loop-observables-cHYM}) in the cHYM  theory
(\ref{eq:2dYM-cHYM-notations})  and its sister aYM theory (\ref{eq:cHYM-2d}).
See \cite{Moore:1997dj,Gerasimov:2006zt,Gerasimov:2007ap} for computation of
correlation functions for the $Q$-closed observables $\tr \phi^n$.

Consider the aYM partition function (\ref{eq:cHYM-2d}).
We can try to proceed in two directions. The first one is to try to use the localization
method and relate the theory to some topological theory and computations with $Q$-equivariant
cohomology. Though the Wilson loop operators are not $Q$-closed, we can try to
solve for at least non-intersecting Wilson loops $\{C_1,\dots, C_k\}$ by: (i) finding topological
wave-function $\Psi(\U_1, \dots, \U_k)$ on the boundary of the Riemann surface
with Wilson loops deleted $\Sigma \setminus \{C_1 \cup
  \dots C_k\}$, and (ii) then integrating over the space of holonomies $\{ U_1, \dots, U_k\}$.
For the study of wave-functions in Higgs-Yang-Mills theory see
\cite{Gerasimov:2006zt,Gerasimov:2007ap}.

%%For solution of Wilson loops in YM by abelianization see \myref{2008-05-27 p 11}.

The second approach is to explicitly solve the constraint $\mu_1  = \mu_2 = 0$,
which means that the complexified connection $A_{\BC} = A + i\Phi$ is flat, in the
form
\begin{equation}
  \label{eq:constraint-solved-by-complexified-group}
  A + i \Phi = g_{\BC}^{-1} d g_{\BC},
\end{equation}
where $g_\BC$ takes value in the complexified gauge group $G_{\BC}$.
The gauge transformations for $g$ taking value in the compact gauge group $G$
\begin{equation}
  \label{eq:gauge-transformation-A}
  A + i \Phi \to g^{-1} (A + i \Phi) g + g^{-1} d g
\end{equation}
can be represented by the right multiplications $g_\BC \to g_\BC g$.
Hence the configurational space of the theory is the same as of gauged
WZW model on the coset $G^{\BC}/G$. We shall not proceed these ideas further in this work.

\appendix
\section{Supersymmetry closure\label{se:off-shell susy}}
 
Let $\delta_{\ve}$ be the supersymmetry transformation generated by a conformal Killing spinor
$\ve$.

% Then the square of $\delta_{\ve}$ is computed as follows
% \begin{equation}
%   \delta_{\ve}^2 A_{M} =\delta_{\ve} (\ve \Gamma_{M} \Psi) = \ve \Gamma_{M} ( \frac 1 2 \Gamma^{PQ} \ve F_{PQ} + \frac 1 2 \Gamma^{\mu A} \Phi_{A} D_{\mu} \ve ).
% \end{equation}
% Since
% \[ \ve \Gamma_{M} \Gamma_{PQ} \ve = \ve \Gamma_{PQ}^T \Gamma_M \ve = -\ve \tilde
% \Gamma_{PQ} \Gamma_{M} \ve = \frac 1 2 \ve (\Gamma_{M} \Gamma_{PQ} - \tilde
% \Gamma_{PQ} \Gamma_{M}) = 2 g_{M[P} \ve \Gamma_{Q]} \ve ,\] the first term for
% $\delta_{\ve}^2 A_{M}$ gives $ -\ve \Gamma^{N} \ve F_{NM}$.  The second term is
% \[ \frac 1 2 \ve \Gamma_{M} \Gamma^{\mu A } \Phi_{A} D_{\mu} \ve = -2 \ve
% \Gamma_M \tilde \Gamma_{A} \ve \Phi^{A}. \] Then
% \begin{equation}
%   \label{eq:delta2onA}
%   \delta_{\ve}^2 A_{M} = -(\ve \Gamma^{N} \ve) F_{NM} -
%   2 \ve \Gamma_{M} \tilde \Gamma_{A} \ve \Phi^{A}.
% \end{equation}
The $\delta_{\ve}^2$ is represented on the fields as  
\begin{equation}
  \begin{aligned}
    \label{eq:delta2onAPhi}
    & \delta_{\ve}^2 A_{\mu} = - v^{\nu} F_{\nu \mu} - [v^{B} \Phi_B, D_{\mu}] \\
    & \delta_{\ve}^2 \Phi_A = - v^{\nu} D_{\nu} \Phi_{A} - [v^{B} \Phi_B, \Phi_A] - 2
    \ve \tilde \Gamma_{AB} \tilde \ve \Phi^{B} - 2\ve \tilde \ve \Phi_A,
  \end{aligned}
\end{equation}
where we introduced the vector field $v$
\begin{equation}\label{eq:v-in-terms-eps-2}
  v^\mu \equiv \ve \Gamma^{\mu} \ve, \quad v^{A} \equiv \ve \Gamma^{A} \ve.
\end{equation}
Therefore
\begin{equation}
  \delta_{\ve}^2 = -L_{v} - G_{v^M A_{M}} - R - \Omega.
\end{equation}
Here $L_{v}$ is the Lie derivative in the direction of the vector field
$v^{\mu}$.  The transformation $G_{v^{M}A_{M}}$ is the gauge transformation
generated by the parameter $v^{M}A_{M}$.  On matter fields $G$ acts as $G_{u}
\cdot \Phi \equiv [u, \Phi]$, on gauge fields $G$ acts as $G_{u} \cdot A_{\mu} =
- D_{\mu} u$.  The transformation $R$ is the rotation of the scalar fields $(R
\cdot \Phi)_{A} = R_{AB} \Phi^{B}$ with the generator $R_{AB} = 2 \ve \tilde
\Gamma_{AB} \tilde \ve$. Finally, the transformation $\Omega$ is the dilation
transformation with the parameter $2(\ve \tilde \ve)$.

On fermions the $\delta_{\ve}^2$ acts as
\begin{equation}\label{eq:delta4fermions}
  \delta_{\ve}^{2} \Psi = -(\ve \Gamma^{N} \ve) D_{N} \Psi -
  \frac 1 2 (\tilde \ve \Gamma_{\mu \nu} \ve) \Gamma^{\mu \nu} \Psi
  - \frac 1 2 (\ve \tilde \Gamma_{AB} \tilde  \ve) \Gamma^{AB} \Psi -
  3(\tilde \ve \ve) \Psi + \text{eom}[\Psi].
\end{equation}

To achieve off-shell closure in the $\CalN=4$ case we add seven auxiliary
fields $K_i$ with $i=1, \dots, 7$ and modify the transformations as
\begin{equation}
  \begin{aligned}
    &  \delta_{\ve} \Psi = \frac 1 2 \Gamma^{MN} F_{MN} + \frac 1 2 \Gamma^{\mu A} \Phi_{A} D_{\mu} \ve + K^i \nu_i \\
    & \delta_{\ve} K_i = -\nu_i \Gamma^{M} D_{M} \Psi.
  \end{aligned}
\end{equation}
Here we introduced seven spinors $\nu_i$. They depend on choice of the conformal 
Killing spinor $\ve$ and are required to satisfy the following relations:
\begin{align}
  \label{eq:nu-relations1}
  &\ve \Gamma^M \nu_i = 0 \\
  \label{eq:nu-relations2}
  &\frac 1 2 (\ve \Gamma_N \ve) \tilde \Gamma^{N}_{\alpha \beta} =
  \nu^i_{\alpha} \nu^i_{\beta} + \ve_{\alpha} \ve_{\beta} \\
  \label{eq:nu-relations3}
  &\nu_i \Gamma^M \nu_j = \delta_{ij} \ve \Gamma_M \ve.
\end{align}
The equation~\eqref{eq:nu-relations1} ensures 
closure on $A_M$, the equation~\eqref{eq:nu-relations2} ensures closure on $\Psi$. 

%These are the key relations which ensure the closure of a single supersymmetry
%transformation generated by $\ve$ even in the $\CalN=4$ case when the set of
%antisymmetric matrices $\Lambda_i$ commuting with $\Gamma^N$ and
%satisfying~\eqref{eq:Lambda-relations} does not exist. 
% The new term in the transformations for $\Psi$ modifies the last line of~\eqref{eq:delta-fermion-on-shell}
% as 
% \[ \delta_{\ve} (K^i \nu_i ) = -(\nu_i \Dslash \Psi) \nu_i. \] 
% Then the terms in $\delta_{\ve}^2
% \Psi$ which were not taken into an account in~\eqref{eq:remain} are
% \begin{equation}
%   -(\nu_i \Dslash \Psi) \nu_i + \frac 1 2 (\ve \Gamma_N \ve) \tilde \Gamma^{N} \Dslash \Psi -
%   (\ve \Dslash \Psi) \ve.
% \end{equation}
% This expression is identically zero because of (\ref{eq:nu-relations2}).  
After adding the auxiliary fields $K_i$, the term proportional to the equations of motion 
of the fermions in~\eqref{eq:delta4fermions} is cancelled and the algebra is closed off-shell.

For the transformation $\delta_{\ve}^2 K_i$ we get
% \begin{equation}
%   \delta_{\ve}^2  K_i = -\nu_i \Gamma^{M} [ (\ve \Gamma_M \Psi), \Psi ]
%   - \nu_i \Gamma^{M} D_{M} (\frac 1 2 \Gamma^{PQ} F_{PQ} \ve + \frac 1 2 \Gamma^{\mu A}
%   \Phi_A D_{\mu} \ve + K^{i} \nu_i).
% \end{equation}

% Using the gamma matrix ``triality identity'' the first term is transformed to
% $\frac 1 2 (\nu_i \Gamma^M \ve) [(\Psi, \Gamma^{M} \Psi)]$, which vanishes
% because of~\eqref{eq:nu-relations1}.  The second term with derivative acting on
% $F$ is equal by Bianchi identity to $(\nu_i \Gamma_N \ve) D_{M} F^{MN}$ and
% vanishes because of~\eqref{eq:nu-relations1}. Then we
% use~\eqref{eq:gamma-tri-spec} to simplify the remaining terms
\begin{equation}
  \delta_{\ve}^2 K_i 
% = - \frac 1 2 \nu_i \Gamma^{\mu} \Gamma^{PQ} \Gamma_{\mu} \tilde
%   \ve F_{PQ} - \frac 1 2 (\nu_i \Gamma^{M} \Gamma_{\mu A} \Gamma^{\mu} \tilde
%   \ve) D_{M} \Phi_A - \frac 1 2 (-\frac 1 {4 r^2}) \Phi_A \nu_i \Gamma^{\nu}
%   \Gamma^{\mu A}
%   \Gamma_{\mu} \Gamma_{\nu} \ve -\\
%   - \nu_i \Gamma^{M} (D_{M} K^{j}) \nu_j- (\nu_i \Gamma^{\mu} D_{\mu} \nu_j)
%   K^{j} = -\frac 1 2 (4) \nu_i \tilde \Gamma^{M B} \tilde \ve D_{M} \Phi_B
%   -\frac 1 2 (-4) \nu_i \tilde \Gamma^{M B} \tilde \ve D_{M} \Phi_B
%   + (\frac 2 {r^2}) \nu_i \Gamma^{A} \ve \Phi_A + \\
%   - (\nu_i \Gamma^{M} \nu_j) D_{M} K^{j} - (\nu_i \Gamma^{\mu} D_{\mu} \nu_j)
%   K^{j} =
%   - (\ve \Gamma^{M} \ve) D_{M} K^{j} -  (\nu_i \Gamma^{M} D_{M} \nu_j) K^{j} =\\
  = - (\ve \Gamma^{M} \ve) D_{M} K^{i} - (\nu_{[i} \Gamma^{\mu} D_{\mu}
  \nu_{j]}) K^{j} - 4(\tilde \ve \ve) K_{i}.
\end{equation}

\bibliography{bsample}
\end{document}